# A new study of the chemical structure of the Horsehead nebula: the influence of grain-surface chemistry

R. Le Gal[1], E. Herbst[1], G. Dufour[2,3], P. Gratier[2], M. Ruaud[2,4], T. H. G. Vidal[2], V. Wakelam[2]

[1] Departments of Chemistry and Astronomy, University of Virginia, McCormick Road, Charlottesville, VA 22904, USA, e-mail: romane.legal@virginia.edu
[2] Laboratoire d'Astrophysique de Bordeaux, Univ. Bordeaux, CNRS, B18N, allee Geoffroy Saint-Hilaire, 33615 Pessac, France
[3] NASA Goddard Space Flight Center (GSFC), Greenbelt, MD, USA
[4] NASA Ames Research Center, Moffett Field, CA, USA

**ABSTRACT**

A wide variety of molecules have recently been detected in the Horsehead nebula photodissociation region (PDR) suggesting that: *(i)* gas-phase and grain chemistries should both contribute to the formation of organic molecules, and *(ii)* far-ultraviolet (FUV) photodesorption may explain the release into the gas phase of grain surface species. In order to tackle these specific problems and more generally in order to better constrain the chemical structure of these types of environments we present a study of the Horsehead nebula gas-grain chemistry. To do so we used the 1D astrochemical gas-grain code Nautilus with an appropriate physical structure computed with the Meudon PDR Code and compared our modeled outcomes with published observations and with previously modeled results when available. The use of a large set of chemical reactions coupled with the time-dependent code Nautilus allows us to reproduce most of the observations well, including those of the first detections in a PDR of the organic molecules HCOOH, $CH_2CO$, $CH_3CHO$ and $CH_3CCH$, which are mostly associated with hot cores. We also provide some abundance predictions for other molecules of interest. Understanding the chemistry behind the detection of these organic molecules is crucial to better constrain the environments these molecules can probe.

## 1. Introduction

The spectral line survey WHISPER[1] (*Wideband High-resolution Iram-30m Survey at two Positions with Emir Receivers*, PI: J. Pety) recently enabled the detection of about thirty molecules up to seven atoms in size (plus their isotopologues) in the Horsehead nebula, a photodissociation region (PDR) situated at the west extremity of the L1630 molecular cloud and illuminated by the $\sigma$-Orionis star (O9.5V) (Gerin et al. 2009). At a distance d $\approx$ 400 pc (Anthony-Twarog 1982), this nearby PDR, seen "edge-on", constitutes one of the brightest filaments in the mid-infrared detected in our galaxy (Abergel et al. 2003).

Given its geometry and intensity, the Horsehead nebula represents an ideal opportunity to study the physical and chemical structure of a PDR. For this purpose, this region has been studied deeply since 2001, in particular with the IRAM-30 meter radiotelescope and the Plateau de Bure interferometer (PdBI). Three different positions have been observed: *(i)* "the PDR position", corresponding to the peak of the HCO line emission, typical of the warmer UV-illuminated gas situated at the top edge of the nebula with a visual extinction of $\sim$ 2 mag (Gerin et al. 2009) ; *(ii)* "the Core", a cold condensation shielded from the UV field, corresponding to the $DCO^+$ line emission, located just after the PDR edge with a much higher visual extinction ($A_V \sim 10-20$ mag) (Pety et al. 2007); and *(iii)* "the PAH position", observed with complementary observations performed with the PdBI (Guzman et al. 2015), closer to the edge of the PDR ($A_V \sim 0.05$ mag), corresponding to the peak at 7.7 $\mu$m of polycyclic aromatic hydrocarbon emission (Abergel et al. 2003).

These three positions correspond to distinguishable regions of the nebula characterized by their different physical conditions (*i.e.* temperature, density, UV penetration) which vary from one region to the other as function of $A_V$. It is therefore of utmost importance to have some constraints on the physical conditions prevailing in each region since they strongly affect the chemistry. The density in the nebula has been determined by Habart et al. (2005) to vary from $\sim 10^2\,cm^{-3}$ in the UV-illuminated outer edges of the PDR and reaching $\sim 2\times10^5\,cm^{-3}$ in less than $10''$(0.02 pc) toward the denser region. These authors also provided a temperature profile as function of the visual extinction via thermal balance modeling. Afterwards, Pety et al. (2007) constrained the gas temperature in the Core region, where the density is $\sim 10^5\,cm^{-3}$, to be $\sim$ 20 K by the deuterium fractionation ratio $DCO^+/HCO^+$. This was confirmed by following studies of the Horsehead nebula, including the WHISPER survey, which enabled the determination of average densities and temperatures to be respectively $n_H = 6\times10^4\,cm^{-3}$ and $T_{gas} = 60$ K for the PDR position, and $n_H = 10^5\,cm^{-3}$ and $T_{gas} = 25$ K for the Core (Gerin et al. 2009; Guzman et al. 2013; Gratier et al. 2013). The dust grain temperature profile has been found to stay relatively low, from $\sim$ 12 K in the Core to almost 30 K at the outer layers of the PDR (Goicoechea et al. 2009a). This low dust temperature profile can be explained as a result of the combination of the moderately low UV-photon radiation field of $60\times$ that of the ISRF (Mathis et al. 1983; Habart et al. 2005; Rimmer et al. 2012) impinging the nebula and the high densities prevailing ($\sim 10^4 - 10^5\,\mathrm{cm}^{-3}$).

[1] http://www.iram-institute.org/~horsehead/Horsehead_Nebula/WHISPER.html

Altogether, the approximately thirty molecules recently detected have confirmed the chemical complexity in the nebula, and generated our interest for the present study. Of these species, we will focus our attention on the seventeen species listed by molecular families in Table 1. As can be seen in this table, the WHISPER survey allowed the detection of some organic molecules in the Horsehead nebula, such as formaldehyde ($H_2CO$) and methanol ($CH_3OH$), which constitute key species in the likely synthesis of more complex organic molecules such as some prebiotic molecules (Bernstein et al. 2002; Muñoz Caro et al. 2002; Garrod et al. 2008). Because they are detected in a wide variety of interstellar sources - in hot cores (Sutton et al. 1995; Ceccarelli et al. 2000), dark clouds (Bergman et al. 2011), shocked regions (*e.g.* Sakai et al. 2012; Codella et al. 2012; Tafalla et al. 2010) and even in comets (Mumma & Charnley 2011; Cordiner et al. 2015) - it is of prime importance to understand well how these precursor molecules form. $H_2CO$ is commonly thought to form both in the gas-phase and on grain surfaces, while $CH_3OH$ is believed to be only formed on grain surfaces (Garrod et al. 2006; Geppert et al. 2006). Guzman et al. (2013) reported the observations of these two molecules toward the Horsehead nebula in both the PDR and Core positions. Unable to reproduce the observed abundances of either $H_2CO$ or $CH_3OH$ at the PDR position with only pure gas-phase models, they concluded that, for this region, both species are formed on grain surfaces and then photodesorbed into the gas phase. On the other hand, at the Core position, a pure gas-phase model can reproduce the observed $H_2CO$ abundance, while photodesorption of ices is still needed to explain the observed abundance of $CH_3OH$. Other organic molecules were reported in the Horsehead nebula as first detections in a PDR environment, including HCOOH (formic acid), CH2CO (ketene), CH3CHO (acetaldehyde), and CH3CCH (propyne) (Guzman et al. 2014). Their abundances were found to be higher at the PDR position than at the Core, revealing that complex organic chemistry is also occurring in UV-illuminated neutral gas (Guzman et al. 2014). Of these molecules, some - HCOOH, $CH_2CO$, and $CH_3CHO$ - have now also been detected in the Orion bar PDR (Cuadrado et al. 2016, 2017).

Nitriles have also been observed in the WHISPER survey, such as acetonitrile ($CH_3CN$), cyanoacetylene ($HC_3N$), and cyanoethynyl ($C_3N$). These species are indeed commonly observed in the interstellar medium, including star formation regions (Bottinelli et al. 2004; Araya et al. 2005; Purcell et al. 2006) and protoplanetary disks (Öberg et al. 2015). The Horsehead $CH_3CN$ spectral lines have been found to be $\sim 40$ times brighter at the PDR position than at the denser Core one (Gratier et al. 2013), suggesting that a surface desorption process, such as photodesorption, should be efficient enough to release organic molecules into the gas phase in far-UV illuminated regions. On the contrary, for $HC_3N$, Gratier et al. (2013) have found more intensive lines in the dense Core than in the PDR region.

Simple hydrocarbons such as CCH, $l{-}C_3H$, $c{-}C_3H$, $c{-}C_3H_2$, and $l{-}C_3H_2$ have also been observed in the WHISPER survey, including the first detection in the ISM of the cation $l{-}C_3H^+$ (Pety et al. 2012). Pety et al. (2012) pointed out that the high abundances found in the Horsehead nebula for these hydrocarbons cannot be reproduced by current pure gas-phase models (Fuente et al. 2003; Teyssier et al. 2004), and suggested a new mechanism of "photo-erosion" by UV radiation of the PAHs to produce them (Pety et al. 2005).

Rotational lines of the carbon fluorine cation ($CF^+$) have been detected with a high signal-to-noise ratio toward both the PDR and the Core positions of the Horsehead (Guzman et al. 2012). Synthesized from the reaction between hydrogen fluorine (HF) and $C^+$ (Neufeld et al. 2005), this cation represents a unique probe, so far, for the fluorine elemental abundance, assuming that $CF^+$ is mainly formed from the reaction $HF + C^+$ and that most of the fluorine is locked in the form of HF. The models of Guzman et al. (2012) predicted that $CF^+$ contains 4-8% of all the fluorine, thus being the second reservoir of fluorine after HF. $CF^+$, observed where $C^+$ is abundant, is also interesting because it can be observed from the ground, while submillimeter satellites (such as *Herschel*) are required to observe $C^+$. Then $CF^+$ could also probe $C^+$, itself probing the cooling gas with its fine structure lines at 157.8 $\mu$m.

With the present study, we aim to understand the chemical content and distribution of the Horsehead nebula according to these recent published observations from the WHISPER survey with a pseudo-time dependent astrochemical model. As far as we know, our model is the first to consider a time-dependent study of the gas-grain chemistry for such environments.

In Sect. 2, we present the astrochemical model developed for this study. In Sect. 3, we compare the time-dependent model results with the observations and analyze for each group of molecules their simulated spatial dependence and the chemical and physical processes responsible for the physical conditions at which our model best reproduces them. Section 4 contains a discussion of our results as well as a comparison with previous studies of the Horsehead nebula chemistry. We also present some predictions from our new model regarding ice abundances and other gas-phase molecular abundances expected in the Horsehead nebula. Finally we summarize our results and draw our main conclusions in Sect. 5.

## 2. A model of the physical and chemical structure of the Horsehead nebula

We modeled the physical and chemical structures of the Horsehead nebula by using a 1D astrochemical model. To do so, we first ran the Meudon PDR Code (Le Bourlot et al. 1993; Le Petit et al. 2006; Le Bourlot et al. 2012) to obtain an appropriate physical structure. Then we used this structure as an input parameter in the 1D pseudo-time-dependent Nautilus code (Ruaud et al. 2016) to compute the chemical time evolution.

### 2.1. Physical structure

The 1D astrochemical Meudon PDR Code is based on a stationary plane-parallel geometry of gas and dust illuminated by an ultraviolet (UV) radiation field coming from one or both sides of the cloud. From starting physical and chemical parameters, the code resolves at each point of the cloud the UV radiative transfer and computes the thermal balance at steady state. We used the Meudon PDR Code to compute the temperature and density profiles of the physical structure of the Horsehead nebula by assuming a cloud illuminated on one side with a moderate UV-photon flux of 60 $\times$ that of the ISRF (Mathis et al. 1983; Habart et al. 2005; Rimmer et al. 2012) and a fixed pressure of $4\times10^6$ K cm$^{-3}$(Abergel et al. 2003) for the PDR part only since for the dense part we fixed the density to $2\times10^5$ cm$^{-3}$. The results are shown in Fig. 1, which displays as functions of the visual extinction from right to left the distance from the UV source and the density profile in the first panel, the molecular hydrogen fraction $f(H_2)$ profile computed by Nautilus and the chemical portion of the Meudon PDR Code in the second panel, and the grain- and gas-temperature profiles in the third panel. The range of dust temperature given by the Meudon PDR Code for different grain size values in the range 0.03 to 0.3 $\mu$m is indicated on the figure. From the third panel in Fig. 1, it can be seen that the temperature of the gas is lower, by a few Kelvin, than the temperature of the dust grains for $A_V \gtrsim 3.5$ mag. This point is discussed in Sect. 4.6. At $A_V \sim 3.5$ mag the gas temperature equals the dust temperature of $\sim$ 12 K, and for $A_V \lesssim 3.5$ mag the gas temperature increases drastically with decreasing visual extinction, reaching $\sim$ 40 K at $A_V \sim 1$ mag, and $\sim$ 130 K at $A_V \sim 0.1$ mag, while the dust temperature only slowly increases, reaching $\sim$ 18 K at $A_V \sim 0.1$ mag. This figure highlights that at the edges of the PDR, dust grains are cooler than the gas, as expected, and already discussed in Sect. 1.

### 2.2. Chemical evolution

Once an appropriate physical structure for the Horsehead nebula was obtained with the Meudon PDR Code, we implemented it in the 1D Nautilus code to compute the time-dependent chemistry. Nautilus is a pseudo-time dependent code based on the two-phase gas-grain model of Hasegawa et al. (1992). First developed to model dense cold gas chemistry, it has recently been extended to a three-phase model by Ruaud et al. (2016), in which the surface and bulk mantle ice phases of a grain are distinguished. For this study we used this most recent version of Nautilus. From an input chemical network, the code builds a system of kinetic rate equations and solves them to obtain the abundances of the species as functions of time. The kinetic equations that determine the evolution of the density of each species in each of the phases are shown in Ruaud et al. (2016).

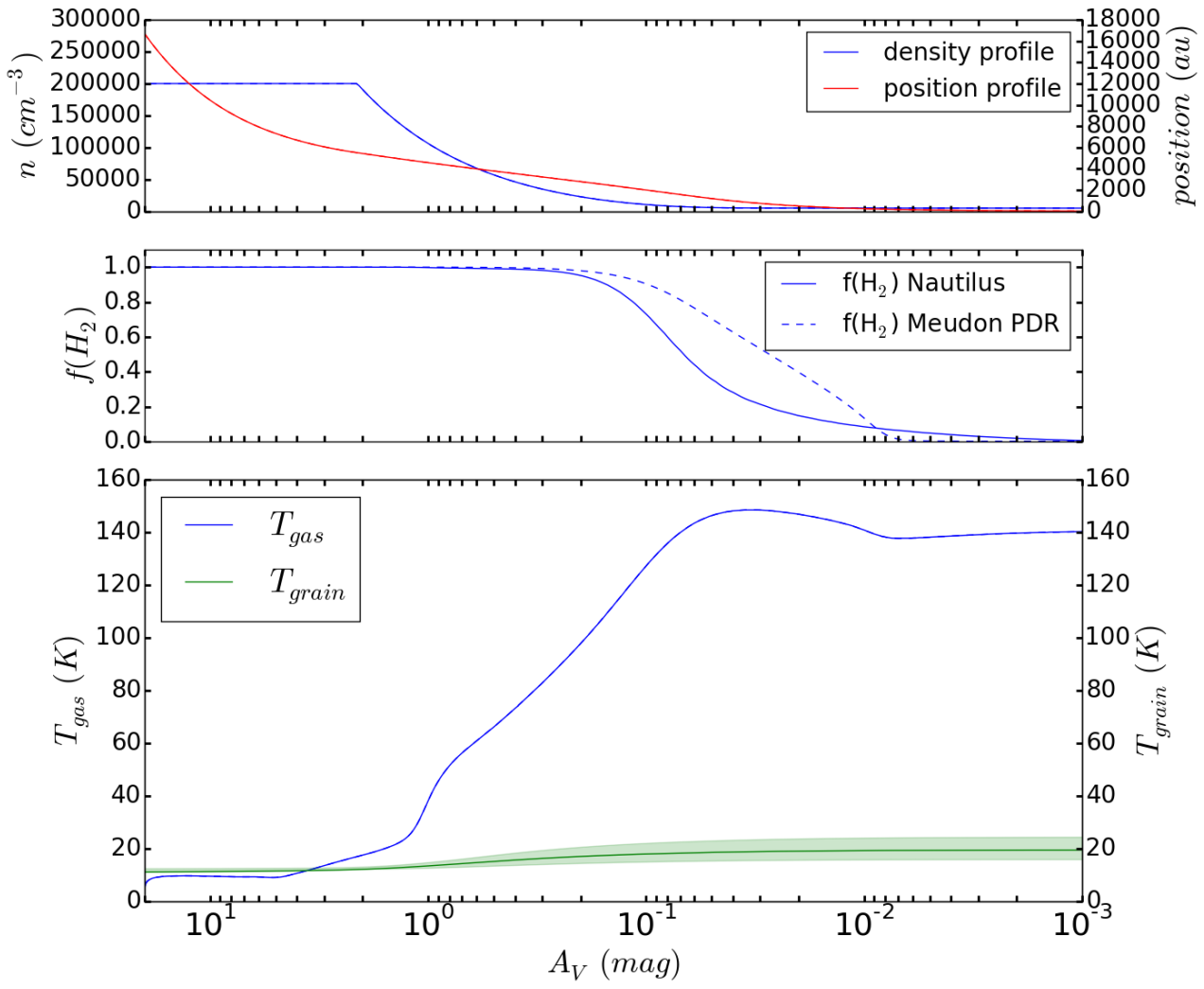

**Fig. 1.** Physical structure of the Horsehead nebula with the density fixed for the dense part of the PDR. All panels are represented as functions of the visual extinction, $A_V$. The first panel shows the density and position profiles, while the second panel represents the molecular hydrogen fraction $f(H_2)$ computed by both codes, with solid line indicating the Nautilus code and dashed line the Meudon PDR Code. The third panel represents the grain- and gas-temperature profiles. The green area covers the range of dust temperature given by the Meudon PDR Code for different grain size values in the range 0.03 to 0.3 $\mu$m.

#### 2.2.1. Chemical processes

In addition to the standard elements of a gas-grain code, the latest version of Nautilus takes into account swapping processes between molecules in ice mantle surfaces and bulk ice mantles (Ruaud et al. 2016). The desorption processes are both thermal, which are inefficient at cold dense cloud temperatures for all but the lightest species, and non-thermal, which includes desorption induced by cosmic-rays (Hasegawa & Herbst 1993), photodesorption, and chemical desorption (Garrod et al. 2007), due to the exothermicity of reactions whith a single product occurring at the surface of the grains. In particular, we assume that, for exothermic reactions leading to one product, 1% of the products desorb. The binding energies considered in this work can be found in table 2 of Wakelam et al. (2017) and on the KIDA database website (http://kida.obs.u-bordeaux1.fr/species.html).

Most of the grain-surface reactions taken into account proceed through the diffusive Langmuir-Hinshelwood mechanism. Even though in the current version of Nautilus (Ruaud et al. 2016) some reactions can proceed through the Eley-Rideal mechanism (Ruaud et al. 2015), we did not include this mechanism. The Nautilus version used here comprises recent updates, such as the reaction-diffusion competition process added by Ruaud et al. (2016). This process allows reactions with chemical activation energy to have this activation energy effectively reduced, and so allows faster diffusion reactions for those with chemical barriers.

Regarding the photodesorption processes, several experimental studies have been done during the last decade for some species of particular interest: CO (Öberg et al. 2007; Muñoz Caro et al. 2010; Fayolle et al. 2011; Bertin et al. 2012, 2013), $H_2O$ (Öberg et al. 2009a), $N_2$ and $O_2$ (Fayolle et al. 2013), $N_2$ and $CO_2$ (Öberg et al. 2009b) and $CH_3OH$ (Öberg et al. 2009a). Two types of processes depending upon whether the excited electronic state is discrete or continuous seem to emerge. But even though experimental coefficients (species desorbed per photon) have been measured, the global processes are still poorly understood (Ruaud et al. 2016), and moreover these processes are highly wavelength dependent (Fayolle et al. 2011; Bertin et al. 2012, 2013). Therefore we chose to not distinguish special cases to be more consistent with respect to the unstudied species, and we followed the recommendations of Bertin et al. (2013) by setting a constant coefficient to $1 \times 10^{-4}$ for all the photodesorptions (Ruaud et al. 2016).

#### 2.2.2. The chemical network

For this study, we used an updated version of the kida.uva.2014 network (Wakelam et al. 2015) specifically including sulfur chemistry updates (Vidal et al. 2017) and recent updates concerning the chemistry of carbon-bearing species such as HCCO, $H_2C_3O$, and the $C_3H_x$ (Wakelam et al. 2015; Loison et al. 2016; Hickson et al. 2016; Hickson et al. 2016). We modified the rate coefficients for the $H_3^+$ photodissociation using the referenced values from the UMIST database, which are consistent with the upper value ($< 1 \times 10^{-12}$ s$^{-1}$) given in van Dishoeck (1987). The chemical network involves 1093 species, with 585 gas-phase species, and 254 grain mantle-surface and bulk-mantle species. These species are linked by a total of 12437 reactions including 8581 gas-phase reactions, 2748 grain mantle-surface reactions including physical processes such as the desorption of molecules into the gas phase and their accretion onto dust-grain mantle surfaces, and 1108 bulk-mantle reactions.

**Table 2.** Elemental abundances and initial species used in this work.

| Species | 0D model[(a)] |
|---|---|
| He | 1.00(-1) |
| N | 7.95(-5) |
| O | 3.02(-4) |
| H | 0.80 |
| $H_2$ | 0.10 |
| $C^+$ | 1.38(-4) |
| $S^+$ | 3.50(-6) |
| $Si^+$ | 1.73(-8) |
| $Fe^+$ | 1.70(-9) |
| $Na^+$ | 2.30(-9) |
| $Mg^+$ | 1.00(-8) |
| $P^+$ | 9.33(-10) |
| $Cl^+$ | 1.00(-7) |
| F | 1.80(-8)[(b)] |

**Notes.** Numbers in parentheses are powers of 10. Elemental abundances of the gas and ice mantle with respect to the total proton abundance. The electron abundance is computed internally in the code in order to have a neutral gas. [(a)] Elemental abundances used in Nautilus code from Goicoechea et al. (2006) and used in Goicoechea et al. (2009b); Pety et al. (2012); Guzman et al. (2012, 2015), except for fluorine. [(b)] Value from Snow et al. (2007), as described in Sect. 2.3.1.

### *2.3. Model description*

#### 2.3.1. 0D model

First, we modeled the chemical evolution during a million years of a starless dense cloud with a high visual extinction. The pseudo-time-dependent model was run for constant physical conditions typical of starless clouds with grain and gas temperatures of 10 K, a gas density $n_{\rm H} = n({\rm H}) + 2n({\rm H_2}) = 2\times10^4\,{\rm cm^{-3}}$, and shielded from the UV photons by a visual extinction of 30 mag. The cosmic-ray ionization rate is set to the constant value of $\zeta = 5\times10^{-17}\,{\rm s^{-1}}$, as previously constrained for the Horsehead nebula by Goicoechea et al. (2006, 2009b), in order to compare our results with those already published (Pety et al. 2012; Guzman et al. 2012, 2015). A single grain radius of 0.1 $\mu$m has been adopted to compute the adsorption rates following Hasegawa et al. (1992).

Assuming that the chemistry starts from diffuse cloud conditions, the chemical components are assumed to be initially in atomic form, except for hydrogen, assumed to be already 20% molecular as considered in previous Meudon PDR Code simulations (Pety et al. 2012; Guzman et al. 2012, 2015). The elements with an ionization potential lower than that of hydrogen (13.6 eV) are initially singly ionized, *i.e.* C, S, Si, Fe, Na, Mg, P, and Cl. For the 0D model of a cold dense cloud we considered the elemental abundances listed in Table 2, which are from Goicoechea et al. (2006) except for fluorine, for which we used a more appropriate diffuse elemental abundance of $1.8\times10^{-8}$ (Snow et al. 2007), justified later in the text. We used these initial chemical conditions in order to better compare our results with the previous astrochemical modeling of the Horsehead nebula (Pety et al. 2012; Guzman et al. 2012, 2015).

#### 2.3.2. 1D model

After a period of $10^6$ yr for a typical interstellar molecular cloud to be formed, we assume that an internal star is also formed, which results in an unshielded moderate UV-photon flux of 60 × that of the ISRF, which then irradiates the portion of the cloud to produce a PDR. The chemical evolution of the "proto-PDR" starts from the output abundances of the first starless molecular cloud model, the 0D model. The chemistry evolves once again over a period of $10^6$ yr for each spatial division of the implemented physical structure of the 1D Horsehead nebula obtained from the Meudon PDR Code. Hereafter our standard model refers to the combination of the 0D and 1D models described in this section.

## 3. Comparison with observations of gas-phase molecules

In this section, we compare our standard model results with the observations of the different species listed in Table 1. Figures 2 - 5 show the modeled and observed abundances in the Horsehead nebula vs. visual extinction for the different categories of species listed in Table 1. The observations are represented by the vertical error bars delimiting the uncertainties in abundance. Because the uncertainties in visual extinction are not observationally well constrained, mainly due to the geometry of the Horsehead nebula seen edge-on, the vertical abundance error bars are placed for clarity where our model reproduces the observations, as functions of $A_{\rm V}$, within their error bars and, in addition, within a factor of ten. The modeling results are represented by the solid curves at so-called early and late times, both of which start from the beginning of the PDR phase. The late time is the same $1\times10^6$ yr for the organic molecules (Fig. 2), the nitrogen-bearing species (Fig. 3), the hydrocarbons of interest (Fig. 4), and $CF^+$ (Fig. 5). The early time is a constant $4.64\times10^4$ yr, except for the hydrocarbons, where an earlier time of $4.64\times10^2$ yr is used to bring out what can happen at very early times. In these figures, we distinguish two regions of the nebula already discussed - the Core and PDR regions, and for the hydrocarbons (Fig. 4) we also include the PAH region for values of the visual extinction under 0.1 mag (Guzman et al. 2015). The grayish boxes indicate the approximate $A_{\rm V}$ ranges at which the observations were performed in each region, according to the corresponding published articles on WHISPER survey observations. Table 3, which contains all the molecules listed in Table 1, gives the extinction ranges in the Core and PDR phases for which our calculated abundances lie within the error bars of the observed values at each time represented in Figs. 2 - 5.

### *3.1. Organic molecules and precursors*

For the PDR region, we can see from Fig. 2 that the selected times do not influence the model abundance results significantly. The observed abundances of $H_2CO$ and $CH_3OH$ can be reproduced by the model at $\sim$ 0.9 – 1.4 mag and $\sim$ 0.7 – 0.9 mag respectively. The HCO observed abundance is best reproduced for slightly higher $A_{\rm V}$ $\sim$ 1.8 – 2.1 mag and that of $CH_3CCH$ at the border between the PDR and the Core regions at $A_{\rm V}$ in $\sim$ 2.7 – 3.2 mag. The three other species $H_2CCO$, $CH_3CHO$ and HCOOH are slightly underestimated by our model by at least a factor of approximately three at their theoretical abundance peaks. But if we consider a standard error factor of ten on the model abundances, these last model results are still consistent with the observations.

For the Core region, Fig. 2 shows that the agreement among the observed species and our calculated abundances is at its best at $1\times10^6$ yr (upper panel), with almost all species calculated te be within their observed error bars for some range of extinction. For example, the observed HCO abundance is indeed well reproduced by our model for either $A_{\rm V} \gtrsim 13$ mag or in the range $\sim$ 5.4 – 5.7 mag where it can be noticed that a general drop in abundance occurs for all of the seven molecules studied here.

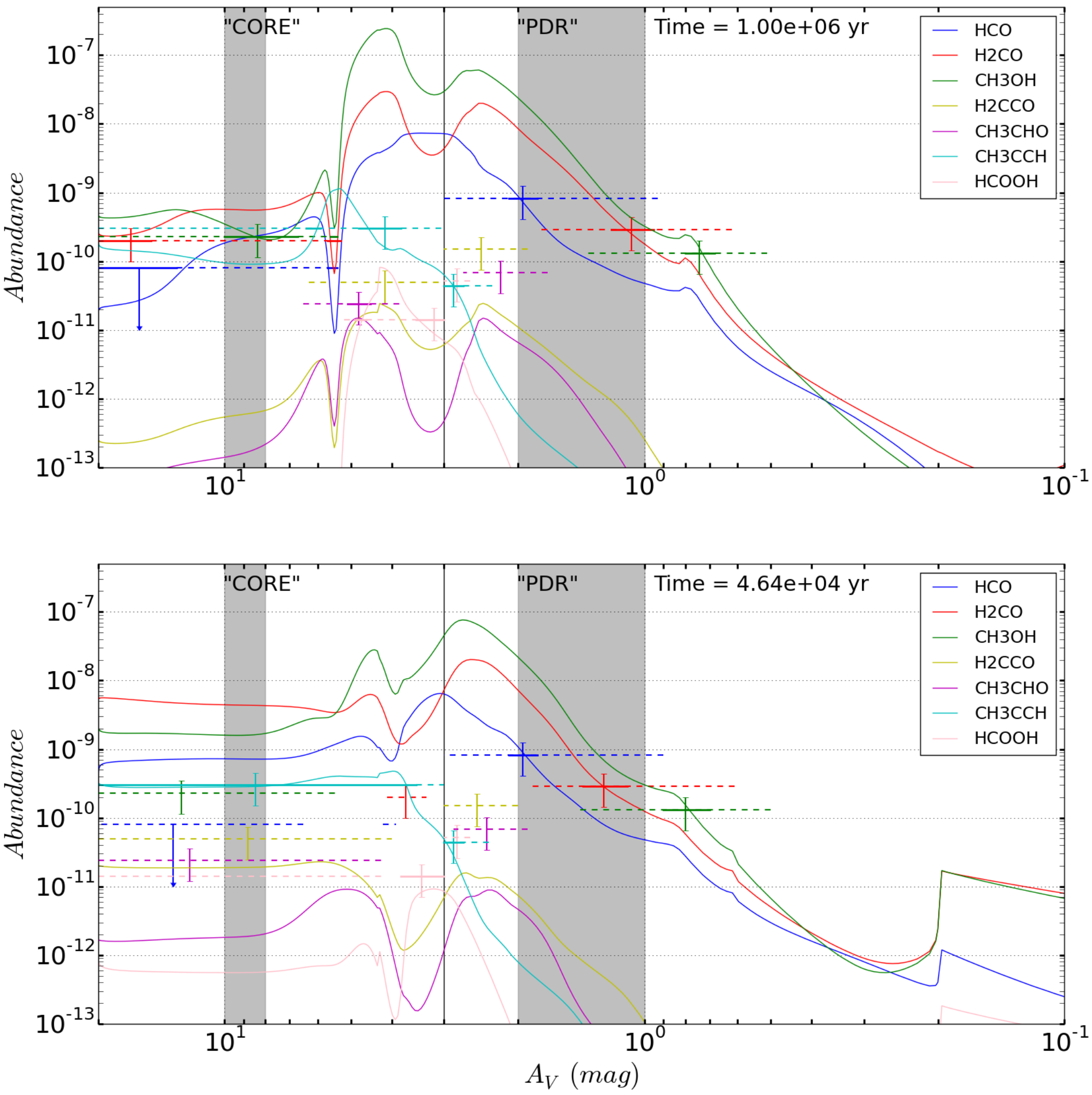

**Fig. 2.** Comparison between the observed and modeled abundances in the Horsehead nebula for HCO and the six organic molecules listed in the first column of Table 1, at two typical early and late times. The model results are represented by solid curves and the observations by vertical error bars which refer to a 50% error in the abundances (Guzman et al. 2014), and horizontal lines, which are solid where our model can reproduce the observed abundances within their error bars and dashed where the agreement is only within a standard factor of ten. The grayish boxes indicate the approximate $A_V$ ranges at which the observations were performed in each region, according to the corresponding published articles.

The ranges of extinction at which the six other species represented Fig. 2 are predicted to lie within their observed error bars are shown in Table 3. At the earlier time of $4.64 \times 10^4$ yr (lower panel), the agreement is not as good in the sense that the abundances of most species are not fit within the shown error bars except for $CH_3CCH$ and, over a very small range of extinctions around its peak calculated abundance, HCOOH. For the other species only a "second order" agreement is obtained by considering a standard factor of ten of error on the calculation.

To summarize, our 0D/1D model seems to be able to reproduce the different observed abundances for the organic molecules of interest in the present study both in the Core and in the PDR positions, even if at a lesser extent for the latter. This is in particular the case for the species first detected in such an environment: HCOOH, $CH_2CO$, $CH_3CHO$ and $CH_3CCH$. Moreover our model is in better agreement with the observations of the Core region at the latest time represented, $1 \times 10^6$ yr, promoting the requirement of a longer timescale for the denser region.

### 3.2. Nitriles

In the PDR region, Fig. 3 shows that our model can reproduce the observations of the three N-bearing species at assorted ranges of extinction at both times depicted. It can also be noticed that the $C_3N$ and $HC_3N$ calculated abundances follow similar trends as functions of the visual extension and do not change much from $4.64 \times 10^4$ yr to $1 \times 10^6$ yr. Thus the model for these two species matches the observations at both times for ranges from $\sim$ 1 to $\sim$ 2 mag and $\sim$ 1.8 to $\sim$ 2 mag, respectively. The $CH_3CN$ modeled abundance also seems not to change much from $4.64 \times 10^4$ yr to $1 \times 10^6$ yr but matches the observations for higher $A_V$, $\sim$ 2.6 – 2.7 mag. This $A_V$ range lies slightly higher than the value of $\sim$ 2 mag derived from the observations (Gratier et al. 2013). At $A_V$ = 2 mag, where the modeled abundances for $HC_3N$ and $C_3N$ reproduce the observations, the model underestimates the $CH_3CN$ abundance by about two orders of magnitude.

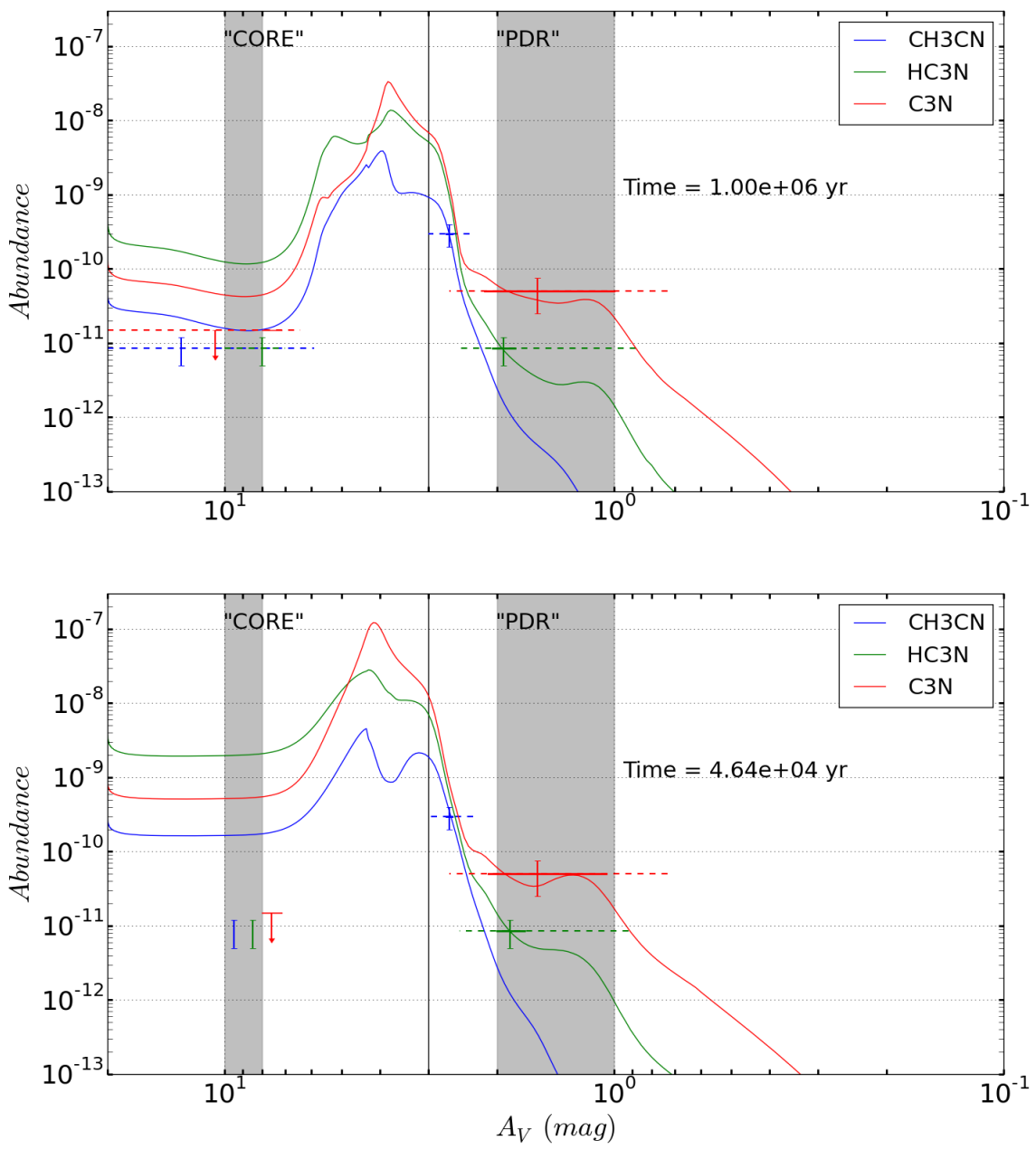

**Fig. 3.** Same as Fig. 2 except that the nitrile abundances are shown and the error bars in the observations come from Gratier et al. (2013).

For the Core region, Fig. 3 shows that the modeled abundances of the three N-bearing species drop by about one order of magnitude for $A_V \gtrsim 7$ mag from $4.64\times10^4$ yr to $1\times10^6$ yr, thus getting closer to the observational error bars at the later time. Nevertheless, the computed abundances still lie outside the observational error bars, but only barely so. At earlier times, the results are significantly worse at high extinction.

One can also note from Fig. 3 that the abundance order, from the most abundant to the least abundant, of the three N-bearing species changes between the PDR and the Core regions: from $C_3N$, $HC_3N$ and $CH_3CN$ in the PDR region to $HC_3N$, $C_3N$ and $CH_3CN$ above $\sim$ 4 mag.

### 3.3. Small hydrocarbons

For the small hydrocarbons, we compare our model results with the observed abundances reported in Guzman et al. (2015), for the Core, PDR and PAH positions. This study took advantage of follow-up interferometric observations performed with the PdBI to further explore and resolve the PDR region layers condensed in a 5″ spatially narrow region (Guzman et al. 2012). The authors have thus been able to better constrain the observation profiles for the small hydrocarbons as functions of the extinction (see Fig. 3 in Guzman et al. 2015). Here, Fig. 4 represents the observed values for the PAH postion, and for the PDR and Core positions where the other species discussed in the present study were also observed within the WHISPER survey. According to the observations of Guzman et al. (2015) the best agreement found with our model seems to be around $A_V$ $1-2$ mag, corresponding to the PDR position. For the PDR region, Fig. 4 shows that depending on the visual extinction, our model seems to reproduce the observed abundances independently from the time, while this is not the case for the Core and PAH regions. Also, another major feature that can be noted from this figure is that the order from the most abundant to the least abundant observed species seems to be pretty well reproduced by our model.

For the PAH region, which lies at the edge of the nebula nearest the exciting star, Fig. 4 displays that our model fails to reproduce the observed abundances. Even at the very early time of $4.64\times10^2$ yr, where the calculated abundances are higher than at $1\times10^6$ yr by several orders of magnitude, the model still underestimates by at least two orders of magnitude the abundances of the three species observed in this region: $C_2H$, c$-C_3H_2$ and $C_3H^+$. One plausible explanation for the discrepancies beteween the observations and our model results is that these small hydrocarbon molecules might be formed from fragmentation of PAHs, as suggested in Pety et al. (2005, 2012).

For the PDR region, one can note that all the neutral three-carbon species considered here can be observed in the same range of extinction, $A_V \sim 0.05$ mag, and that $C_2H$ and the $C_3H^+$ ion have their own specific behaviours differing from one another for $A_V \gtrsim$ 0.5 mag or $A_V \gtrsim 0.9$ mag depending on the time. For $A_V \sim 1-3$ mag, the calculated $C_3H^+$ abundance decreases by more than three orders of magnitude in this visual extinction range at both times. The $C_3H^+$ presence is indeed extremely localized and depends on several parameters fine-tuned together, including the $H_2$ abundance and ortho-to-para ratio, as well as the temperature. This can explain that despite the quantity of sources where $C_3H^+$ has been searched for (McGuire et al. 2014), it has only been detected in the Horsehead nebula (Pety et al. 2012; Guzman et al. 2015) and in the Orion Bar (McGuire et al. 2014; Cuadrado et al. 2015). The temperature dependence is caused mainly by the strong inverse temperature dependence of the rate coefficient of the $C_3H^+ + H_2$ reaction (Savić & Gerlich 2005). As seen in Fig. 1 the gas temperature decreases from $\sim$ 40 to 12 K in the range $A_V \sim 1-3$ mag, which might explain the diminution of the $C_3H^+$ ion in this $A_V$ range.

For the Core region, our model is only in agreement with the upper limit derived for the observations of $C_3H^+$, independently of time. All the other calculated hydrocarbon abundances are overestimated by at least an order of magnitude: a factor of approximately twelve for c$-C_3H$, l$-C_3H$, and l$-C_3H_2$ and approximately two orders of magnitude for c$-C_3H_2$ at the earlier time shown Fig. 4. We can also see that the modeled abundances of l$-C_3H$, c$-C_3H$ and l$-C_3H_2$ increase with time, by factors from two to four, while c$-C_3H_2$ decreases by factors of between two and five, depending on the $A_V$ considered, reflecting the chemical formation and destruction pathways of these molecules discussed below.

### 3.4. Chemistry of unsaturated hydrocarbons

Guzman et al. (2015) discussed the known gas-phase synthetic processes for the small hydrocarbons of interest here, up to $C_3H^+$. This ion can form through different pathways such as $C_2H_2^+$ (C, H) $C_3H^+$, $C_2H_2$ ($C^+$, H) $C_3H^+$, and $C_2H$ ($C^+$, H) $C_3$ followed by $C_3^+$ ($H_2$, H) $C_3H^+$. With our present astrochemical model, the third pathway dominates independently of the visual extinction until a few $10^2$ yr and then the $C_2H_2 + C^+$ pathway dominates up to $10^6$ yr (except for $A_V > 4.3$ mag where the $C_2H_2^+ + C$ also dominates between $\sim 3\times10^2$ and $\sim 4\times10^3$ yr). Meanwhile the predominant destruction pathway of $C_3H^+$ is the recombination with electron independent of the time and the visual extinction. $C_3H^+$ is supposed to be the main precursor for $C_3H$ and $C_3H_2$ via successive hydrogenations with $H_2$ leading to the ions $C_3H_2^+$ and $C_3H_3^+$, followed by electronic dissociative recombinations to form the neutral hydrocarbons (Maluendes et al. 1993). Even if $C_3H_2$ is related to $C_3H^+$ via its formation pathway, the fact that the two species do not have the same spatial distribution does not obviously mean that the gas-phase chemistry alone is insufficient to explain the observations (Guzman et al. 2015). As a counter example, this effect can arise because the destruction pathways differ from one molecule to the other, which might give information on the chemical timescale. Depending on the $A_V$, $C_3H_2$ is mainly destroyed either by photons, or as the extinction become higher ($A_V \gtrsim 1.3$ mag), by $C^+$ for the cyclic form, and by H for the linear form. On the contrary, $C_3H^+$ is mainly destroyed by electrons independently of the extinction, which could explain the discrepancy found between the spatial distributions of $C_3H^+$ and $C_3H_2$. Moreover, Fig. 4 indeed shows that the chemical behaviors of the two species seem to diverge at later times, which might lead to different spatial distributions. Thus the discrepancy found between the $C_3H^+$ and $C_3H_2$ spatial distributions (Guzman et al. 2015) could potentially indicate some new plausible "chemical clock".

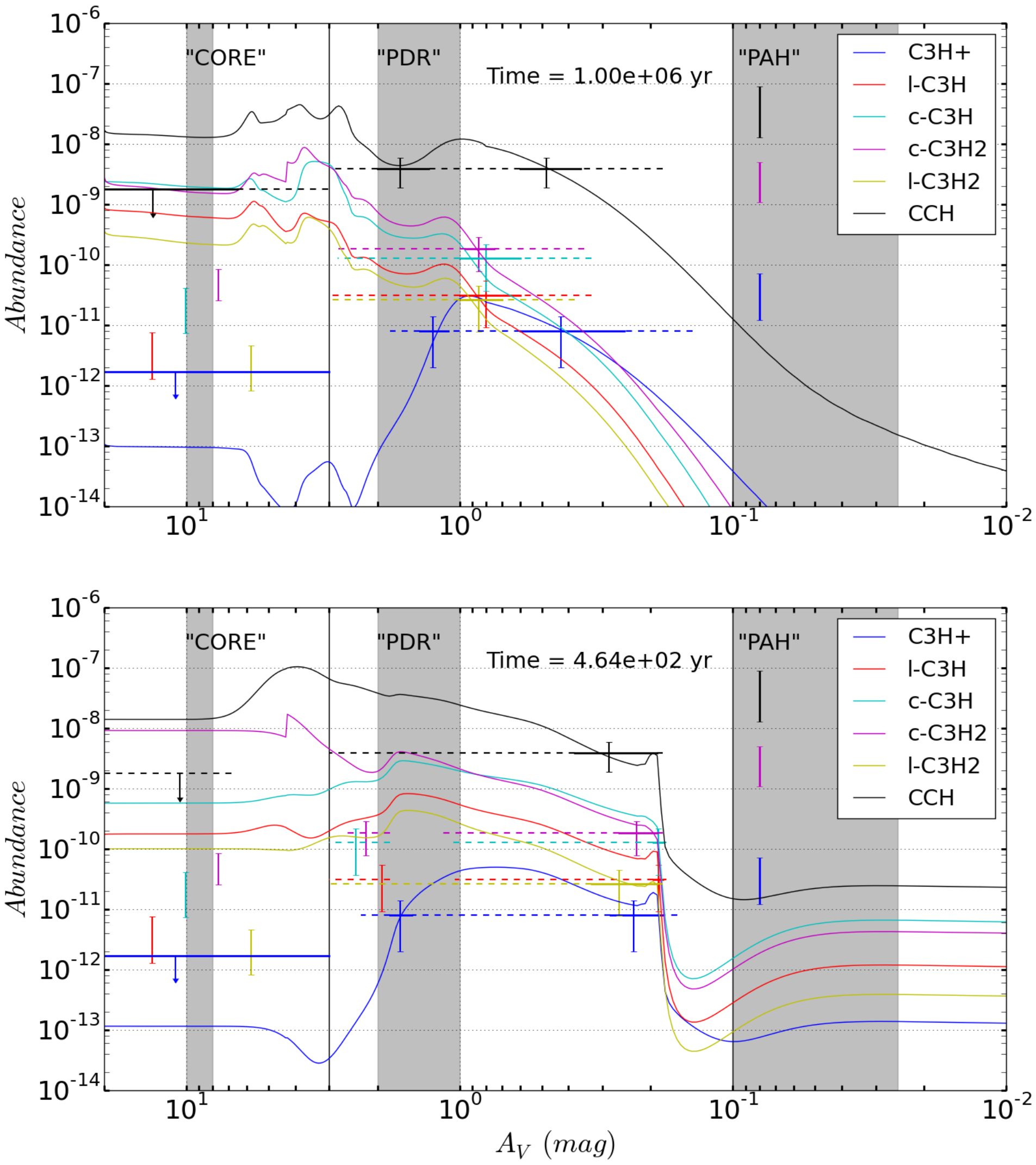

**Fig. 4.** As Figs. 2 and 3 except that the small hydrocarbon abundances are plotted, the early time value is two orders of magnitude lower, the PAH region is included, and the error bars in the observations are from the interferometric study of Guzman et al. (2015).

### 3.5. $CF^+$

In Fig. 5, we compare the observed and calculated abundances of $CF^+$ in both the Core and the PDR positions at $4.64 \times 10^4$ yr and $1 \times 10^6$ yr. Since our standard model, hereafter Model 1, is not able to reproduce the $CF^+$ abundance well, we have run two additional models: Model 2, which considers a lower elemental abundance for fluorine of $6.68 \times 10^{-9}$ (Goicoechea et al. 2006), and Model 3, considering new values for rate coefficients for the following important reactions:

- $CF^+$ + Photon $\longrightarrow$ F + $C^+$ (Guzman et al. 2012),
- HF + $C^+$ $\longrightarrow$ H + $CF^+$ (Neufeld et al. 2005; Guzman et al. 2012),
- $CF^+ + e^-$ $\longrightarrow$ C + F (Novotny et al. 2005; Neufeld & Wolfire 2009; Guzman et al. 2012),
- F + $H_2$ $\longrightarrow$ HF + H (Tizniti et al. 2014).

Figure 5 shows that, for the PDR region, the observations are best reproduced by Model 3 which contains the updated fluorine chemistry and a less depleted fluorine elemental abundance (see Table 2 and Sect. 2.3.1). A comparison between the upper and lower panels of Fig. 5 also shows that the $CF^+$ abundance is independent of time for the PDR region, but, for $A_V > 3$ mag, the $CF^+$ abundance decreases with time.

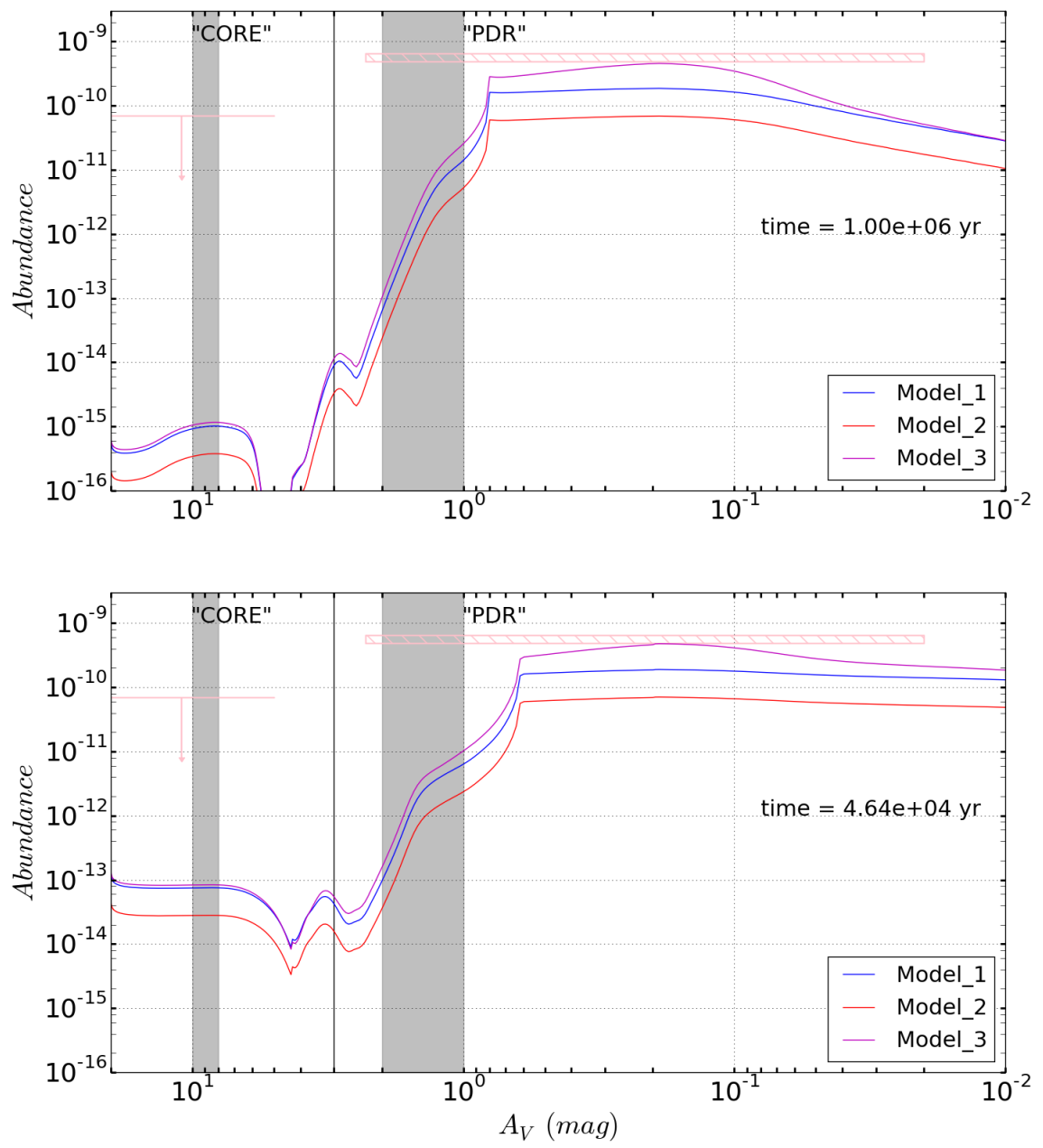

**Fig. 5.** Comparison between the observed Guzman et al. (2012) and calculated abundances of $CF^+$ in both the Core and the PDR positions of the Horsehead nebula at: $4.64 \times 10^4$ yr and $1 \times 10^6$ yr. Three different models are considered here concerning the fluorine chemistry: Model 1, corresponding to the one described in Sect. 2.3; Model 2, which contains a lower fluorine abundance (Goicoechea et al. 2006), and Model 3, with improved reaction rate coefficients discussed in the text. The model results are represented by the solid curves and the observations by the upper limit and hatched box represented in pink.

## 4. Discussion of some calculated abundances and predictions

Overall the calculated results obtained with our 0D/1D model reproduce reasonably the observed abundances derived toward the Horsehead nebula and especially the time-dependent characteristic of our simulation allows us to argue in favor of late chemical timescales to reproduce the Core region abundances. We discuss in this section our results in the light of the recent progress made in understanding the chemistry of the different families of species presented in this study. Quantitative results containing abundances and comparisons with observations are found in Table 4, in which we present theoretical molecular abundances obtained with our standard model at a timescale of $1 \times 10^6$ yr for three specific $A_V$ values corresponding to those derived from the observations of the PAH, PDR, and Core postions (Guzman et al. 2015; Pety et al. 2012; Gerin et al. 2009). Figures 8 - 10 represent the abundances of some non-observed and observed molecules not yet discussed in this study, as functions of the visual extinction for early and late times as done in Figs. 2 - 7 to give a global overview of the behavior of each species with the extinction. These abundance predictions are discussed Sect. 4.5.

### 4.1. Organic molecules

The organic molecules detected wih the WHISPER survey have been found more abundant at the PDR than at the Core position except for methanol for which the reverse is observed (Guzman et al. 2014). Moreover, according to Guzman et al. (2014), methanol is less abundant in the Horsehead nebula than toward hot core sources where it has been mostly detected (Bisschop et al. 2007). On the other hand, other organic molecules detected in the Horsehead nebula have similar abundances to those found in hot cores (Bisschop et al. 2007) and prestellar cores (Bacmann et al. 2012). This interesting result suggests that methanol is probably tracing a hotter gas where thermal desorption or even explosion of grain mantles (Coutens et al. 2017) can occur, while the other detected organic molecules (*i.e.* $CH_2CO$, $CH_3CHO$, HCOOH and $CH_3CCH$) might trace colder regions. In these colder regions, the dominant desorption mechanism proceeds probably mainly by chemical desorption, as with our present model and the model developed by Esplugues et al. (2016), instead of photodesorption as previously suggested (Guzman et al. 2014) due to overestimated photodesorption rates (Bertin et al. 2016; Cruz-Diaz et al. 2016). To study the efficiency of the photodesorption mechanism in releasing the organic molecules into the gas phase with our model, we ran two similar models: one without taking into account photodesorption and another with a photodesorption yield of $1 \times 10^{-3}$. We obtained very similar abundances to those represented and described in Sect. 3.1. Thus, according to our model, the photodesorption is not a main pathway to release molecules in the gas phase. In our preliminary 0D model, the main process that desorbs the molecules is chemical desorption. Subsequently, in our 1D model, depending on $A_V$, the organic molecules are mainly released into the gas phase by chemical desorption for the denser parts and, for the less extinguished regions, also by thermal desorption due to the higher temperatures prevailing. Therefore the fact that the organic molecule abundances, observed in the Horsehead, are higher in the PDR region than in the Core might indicate that grain surface processes are sped up thanks to an enhanced mobility on grains caused by the warming up of their surfaces. This warming could accelerate diffusion processes (Vinogradoff et al. 2013; Mispelaer et al. 2013) and thus increase the production of organic molecules on grain surfaces.

The abundance peaks for most of the complex organic molecules discussed in the present study are found to lie in between the PDR and Core regions. This is indeed expected since at lower densities there is less depletion onto grain surfaces, resulting in fewer grain-surface reactions and thus a lowered production of organic molecules. At higher densities, while more species are depleted onto grains, there is less desorption occuring so there are fewer organic molecules in the gas phase (see Fig. 6). The high peak abundances might also be generated by inclusion of a competition between diffusion and reaction over chemical barriers, occasionally used to analyze experimental data and contained in the model used (Ruaud et al. 2016). Such a competitive process is indeed more realistic and required to better reproduce the gas-phase abundances for species produced by reactions with barriers (Herbst & Millar 2008) since without it methanol is under-estimated by three orders of magnitude (Ruaud et al. 2016).

Moreover, even though Guzman et al. (2014) conclude that the WHISPER survey results seem in agreement with the fact that none of the typical "hot molecules", such as $CH_3OCH_3$ or $HCOOCH_3$, is detected in the Horsehead, we present here some abundance predictions for these complex organic molecules based on their recent surprising observations in cold prestellar core ($T \sim 10$ K) (Bacmann et al. 2012). Figure 6 represents the calculated abundance predictions as functions of the visual extinction obtained with our 0D/1D model for the complex organic molecules (COMs) dimethyl ether ($CH_3OCH_3$) and methyl formate ($HCOOCH_3$), both in the gas phase and on ices, and also for the radical $CH_3O$, one of the main precursors of saturated COMs (Brown et al. 1988; Garrod et al. 2006; Herbst & van Dishoeck 2009). According to our model $CH_3O$ should be detectable in the gas phase with a reasonable fractional abundance of $\gtrsim 10^{-11}$ for $A_V \sim 2-5$ mag. Regarding the two COMs $HCOOCH_3$ and $CH_3OCH_3$, their gas-phase fractional abundances are low for all visual extinction values and only become higher than $\gtrsim 10^{-13}$ for $A_V \sim 2-5$ mag, which is consistent with their non-detection in the Horsehead (Guzman et al. 2014). One can indeed see in Fig. 6 that $CH_3O$ is more abundant in the bulk mantle at both plotted times, and $CH_3OCH_3$ is also more abundant on ices, either inside the mantle or at its surface, while $HCOOCH_3$ is more abundant in the gas-phase until an $A_V$ of $\sim 5$ mag at $4.64 \times 10^4$ yr and $\sim 7$ mag at $1 \times 10^6$ yr. We note that the KIDA chemical network used for the present study does not include yet the new gas-phase COM synthesis recently suggested by Balucani et al. (2015).

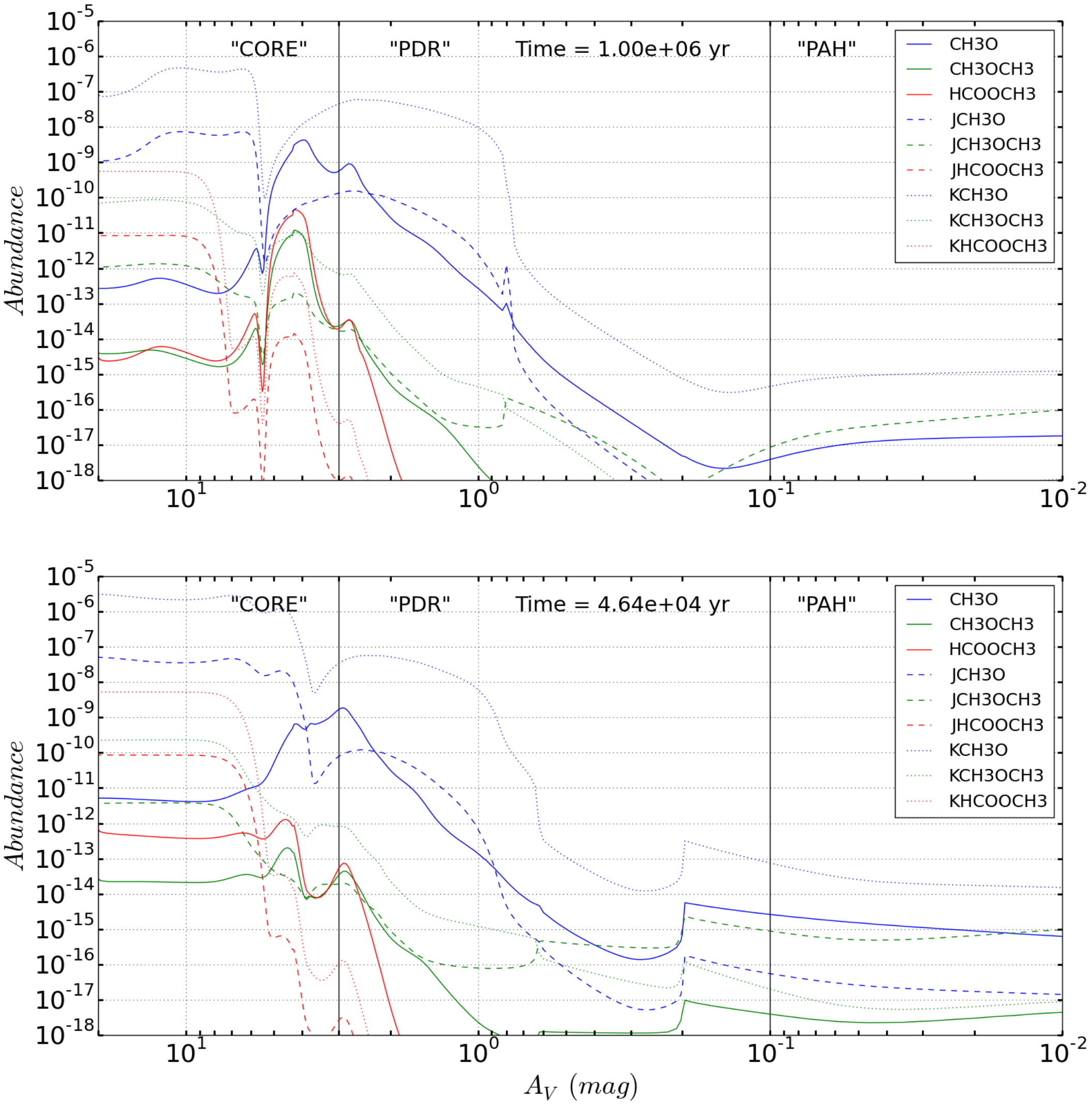

**Fig. 6.** As Figs. 2 - 5 but here representing the predicted gas-phase and ice abundances in the Horsehead nebula for $CH_3O$, $CH_3OCH_3$ and $HCOOCH_3$, at the same typical early and late times: $4.64 \times 10^4$ yr and $1 \times 10^6$ yr. The "Jspecies" designate the mantle-surface molecules and the "Kspecies" the bulk-mantle molecules.

Figure 7 shows the ice abundance predictions for the organic molecules listed in the first column of Table 1 and for which we presented the gas-phase abundances in Fig. 2. These predictions are given for information only since today it might be difficult to observe the ice features in the Horeshead. However, they could be used in the future to indicate which ice molecules could be observable if such observations appear to be possible, as with JWST.

### 4.2. Nitrile case

One fifth of the $\sim 200$ molecules detected up to the present belong to the nitrile (R-CN) / isonitrile (R-NC) family. Acetonitrile ($CH_3CN$) is a commonly observed nitrile in the ISM, and is thought to be a good temperature probe because it is a symmetric rotor presenting metastable levels (Guesten et al. 1985). The fact that $CH_3CN$ was observed with a higher abundance in the PDR than in the Core, as opposed to $HC_3N$, made Gratier et al. (2013) suggest that a surface desorption process, such as photodesorption, should be occurring to release organic molecules into the gas phase in far-UV illuminated regions. These authors developed a simple steady-state gas-phase chemical model which underestimated the $CH_3CN$ abundance in the PDR by approximately three orders of magnitude. Our model also underestimates the $CH_3CN$ abundance by about two orders of magnitude, in the PDR region ($A_V \sim 2$ mag), where the calculated abundances for $HC_3N$ and $C_3N$ reproduce the observations. But our model can reproduce the observed $CH_3CN$ abundance for higher visual extinction suggesting that the species might not be found at the same position. For the Core region, the agreement with the observation is less good but still better for the later time of $1 \times 10^6$ yr, such as obtained for the organic molecules.

So, our model is once again able to reproduce the observations depending on the visual extinction. The high acetonitrile can be explained by a higher photodesorption rate and a lower ice photolysis rate than those currently assumed in the models (Gratier et al. 2013; Ligterink et al. 2015). The Bernstein et al. (2004) results on a slower photolysis process for solid $CH_3CN$ than for other organic molecules argue in this sense as do those of Danger et al. (2011) concerning $CH_3CN$ formation by UV photolysis of ethylamine ($CH_3CH_2NH_2$) in ices. Gratier et al. (2013) also highlighted the fact that the species has also been observed in shocks (Arce et al. 2008; Codella et al. 2009) indicating that acetonitrile might be subject to sputtering in those regions. Moreover, recent experimental studies focused on the specific topic of high-energy irradiation bombardment of $CH_3CN$ ices conclude that this process could enhance the complex nitrile abundances in the gas phase (Ribeiro et al. 2015).

### 4.3. Small hydrocarbon case

The high values found for the small hydrocarbons in the Horsehead nebula PAH region might pinpoint that a "top down" chemistry is occuring in those environments, in which small hydrocarbons form from large molecule fragmentation, such as PAHs, with a high UV flux (Fuente et al. 2003; Pety et al. 2005, 2012; Guzman et al. 2015). Nevertheless, our model is able to reproduce the hydrocarbons observed toward the PDR region and even overproduce these same hydrocarbons for the densest region. However, we note that the observed abundances for the small hydrocarbons are particularly low in the Core region compared with typical dense cloud values such as the case of TMC-1 (Agúndez & Wakelam 2013).

On the other hand it is true that for the PAH region, corresponding to the outside edges of the PDR region, the hydrocarbon abundances are underestimated and might require a new type of formation pathway or might suggest that a higher ionization rate has to be considered for the edges of the PDR. This last point has indeed been explored in Rimmer et al. (2012), who studied the influence of the cosmic-ray ionization rate and showed that high value of $\zeta$ – either a fixed $\zeta \sim 10^{-15}$ s$^{-1}$ or a "high range" column-dependent ionization rate $\zeta(N_H)$ – is in better agreement with observations for both the PDR and the Core regions. On the other hand, in the PAH region the hydrocarbon abundances seem not to depend much on $\zeta$, probably becauce in this region, the hydrocarbon chemistry is dominated by photon interactions than by cosmic-rays'. It should thus be worthwhile to further explore the impact of the cosmic-ray ionization rate in our model to see how it impacts the chemistry and if it helps to reproduce better the small hydrocarbons in the PAH region, even though Rimmer et al. (2012) found with their model that it does not.

### 4.4. $CF^+$ case

In Sect. 3.5 we presented the impact of updates on the $CF^+$ chemistry. We confirmed that indeed the fluorine elemental abundance of $(0.6 - 1.5) \times 10^{-8}$ derived from the $CF^+$ observation in the Horsehead nebula (Guzman et al. 2012), consistent with the diffuse gas value (Snow et al. 2007), allows a better agreement between observations and models than the previous lower value of $6.68 \times 10^{-9}$ (Goicoechea et al. 2006). Moreover the updates concerning the $CF^+$ chemistry modeling also improved the agreement with the observations.

### 4.5. Additional predictions

It would be desirable if our calculated results could be compared with previous time-dependent gas-grain model calculations. There is a previous time-dependent chemical model of the Horsehead nebula by Rimmer et al. (2012) but it is different from ours in three respects: (i) it uses a variable value for the cosmic-ray ionization rate $\zeta$ that diminishes as the cosmic rays enter the nebula from its edge, (ii) it does not contain grain-surface chemistry, and (iii) it appears to utilize incorrect values for the visual extinction in the plane of the nebula, which are significantly larger than used here. So, to the best of our knowledge, the majority of our results can only be compared with observational abundances. However we present here the predictions given by our model for some of the molecules that Rimmer et al. (2012) presented.

As highlighted by Rimmer et al. (2012) it might be worthwhile to look at the HCN and HNC molecules, which according to our respective models should be present with sufficient amount to be observable mostly in the inner regions. As depicted in Fig. 8, our calculated HNC/HCN ratio seems to slightly decrease with increasing visual extinction until $A_V \simeq 3$ mag, from a value of approximately one to a value decreased by slightly more than one order of magnitude at $A_V \simeq 3$ mag. It then reincreases with visual extinction until a HNC/HCN ratio of $\sim 0.8$. Our model also shows that more complex carbon-chain nitriles might also be detected, such as $HC_5N$. Ammonia should also be observable in reasonable amount according to our model. We would like also to emphasize that CN and NO might also be observable with the particular interest of their being potential tracers of the C/O elemental abundance ratio.

Le Gal et al. (2014) indeed showed that the nitrogen chemistry depends highly on the C/O gas-phase elemental ratio. This is mainly because carbon and oxygen are the most abundant elements after hydrogen and helium in the interstellar medium, and are found in similar quantities. Therefore the elemental C/O ratio can vary with values higher or below unity, entailing non negligible consequences on the interstellar chemistry. Moreover, except for the reference value of C/O$\sim 0.55$ derived toward the Sun (Asplund et al. 2009), we still do not know precisely how this ratio varies with environment. Astrochemical models show that the CN/NO ratio can vary between 6 and 10 orders of magnitude for a C/O ranging from 0.3 to 1.5 for typical dense cold gas conditions (Le Gal et al. 2014). The CN/NO ratio can thus a priori appear as an ideal probe for the elemental C/O ratio.

Regarding the oxygen-containing ions $OH^+$, $H_2O^+$ and $H_3O^+$, Fig. 9 shows that, independently of $A_V$, $H_3O^+$ is more abundant than $H_2O^+$, which is itself more abundant than $OH^+$ at early time. At later time, these ionic abundances increase with decreasing visual extinction and in the PAH region for $A_V \lesssim 0.04$ mag, the abundance order is completely reversed, *i.e.* with $OH^+$ more abundant than $H_2O^+$ itself more abundant than $H_3O^+$. Overall, the ion $H_3O^+$ reaches its calculated peak abundance as a function of $A_V$ ($\sim 9 \times 10^{-10}$ at a time of $1 \times 10^6$ yr) at $A_V \sim 4$ mag which seems to be also in good agreement with Rimmer et al. (2012) results who found a higher abundance for $H_3O^+$ at $A_V \simeq 4.5$ mag than at the two other visual extinction values they explored ($A_V \simeq 1.5$ mag and $A_V \simeq 12$ mag). So, as Rimmer et al. (2012), our model predicts that $H_3O^+$ might be observable in the inner regions of the Horsehead. Thus, in opposition to the Rimmer et al. (2012) conclusion about these ions, we predict that all the three ions might be detectable at the extreme edge of the PAH regions.

Concerning the sulfur chemistry, CS, $C^{34}S$ and $HCS^+$ have been observed in the Horsehead nebula (Goicoechea et al. 2006) and seem to indicate that the elemental sulfur depletion of several orders of magnitude is no longer required to explain the abundances of these molecules. Instead, depletion of a factor a few with respect to the solar abundance is sufficient. The Rimmer et al. (2012) modeling study support this assessment with also better reproducing the observed sulfur-bearing species abundances when taking into account a higher elemental sulfur abundance. This is why we ran our models with the "high" sulfur elemental abundance value recommended by these authors of $3.5 \times 10^{-6}$ as shown in Table 2. But in order to verify that our model produce sulfur-bearing molecular abundance consistent with those of Goicoechea et al. (2006) and Rimmer et al. (2012) we also ran a "low" sulfur elemental abundance model, with an initial elemental sulfur abundance of $8.0 \times 10^{-8}$ (Graedel et al. 1982; Wakelam & Herbst 2008). The results of the two different models are compared in Fig. 10 which represents the observed and calculated sulfur-bearing species abundances for the high and low sulfur amount models. It can be seen that both the high and low sulfur elemental abundance models can reproduce the observations and even that for the Core region the low sulfur elemental abundance case reproduces the observations better. The fact that our results seem different from those of Rimmer et al. (2012) is mainly because these authors directly compared their model fractional abundances, relative to the total H nuclei, with the observed values from Goicoechea et al. (2006) which are relative to $H_2$. Due to this inconsistency, we chose to compare our results with the observed values relative to the total H nuclei, as reported in Guzman et al. (2014).

### *4.6. Investigation of the temperature profile*

A notable point in our modeling is that the temperature profile we obtained with the Meudon PDR Code gives a gas temperature lower than the grain temperature in the cloud interior, as represented in Fig. 1 and mentioned in Sect. 2.1. This occurs because in the public version of the Meudon PDR Code we used, the freeze-out of species from the gas phase onto the grain surfaces is not included. Consequently, the gas and grain temperatures are not rigorously self-consistent with the chemistry.

If freeze-out were included in the Meudon PDR Code, it would remove some gas coolants from the gas phase (such as CO) in the cloud interior and thus result in a higher gas temperature for the denser regions. In order to investigate the impact of this change in the temperature profile we ran some additional models for which we changed the temperature profile by setting the gas temperature greater or equal to the dust temperature. The results are not significantly affected, probably because the dust temperature is low ($\approx 12$ K) in the cloud interior, so that even if the gas temperature reaches lower values it does not impact the chemistry significantly.

## 5. Summary and conclusions

In this study, we developed a gas-grain chemistry model with time dependence to explore and understand the rich chemistry occurring in the Horsehead nebula. The initial 0D astrochemical model allows us to build a starless initial molecular cloud to represent the birth place for the star $\sigma$ Ori. We used the code Nautilus to run this 0D model for a period of $10^6$ yr, and then assumed that $\sigma$ Ori has been formed. Afterwards we ran a subsequent 1D astrochemical model to take into account the impact of the FUV-flux coming from the star, which impinges the Horsehead nebula. The initial conditions for the 1D astrochemical model are the output chemical abundances of the 0D model run coupled to a physical structure, including the density and temperature profiles, previously computed with the Meudon PDR Code. This 0D/1D model is to the best of our knowledge the first of its kind and allows us to investigate the chemistry and the chemical timescale impact further. In particular the time dependence of our model allows us to study realistically the denser part of a PDR where the usual steady-state PDR model will not be appropriate (Esplugues et al. 2016). We emphasize two typical early and late times and find that the late-time model better reproduces the observations. The modeling results obtained have shown that our gas-grain time-dependent model seems appropriate for modeling the chemistry of such an environement and provides some predictions for future observational campaigns, which might lead to a deeper understanding of the Horsehead nebula chemistry and elemental composition, such as for instance the elemental C/O ratio. In particular, our model supports the conclusion of Esplugues et al. (2016) that the presence in the gas-phase of $H_2CO$ and $CH_3OH$ and other COMs is mainly due to chemical desorption and not as previously thought to photodesorption (Guzman et al. 2014).

The laboratory work on chemical reactive desorption by Dulieu et al. (2013) and further explored in Minissale & Dulieu (2014), has recently been summarized in Minissale et al. (2016). In this last work, a semi-empirical theory was derived that describes the dependence of the efficiency of chemical reactive desorption on the type and chemical composition of the surface, on the enthalpy of the surface reaction considered, and on the binding energy of the desorbing molecule. This new formula was used in a subsequent modeling study (Wakelam et al. 2017), which, when combined with new binding energy computations, greatly lowers the chemical desorption rate and seems to rule out the reproduction of the cold core COM observations by grain-surface chemistry. It might thus be worthwhile to further explore the impact of the chemical desorption probability for the denser part of PDRs.

*Acknowledgements.* The authors would like to thank the anonymous referee for valuable suggestions and comments. We would like to thank also Jean-Christophe Loison, Evelyne Roueff and Viviana Guzman for useful discussions. EH wishes to thank the National Science Foundation for continuing to support the astrochemistry program at the University of Virginia. VW, MR, PG and GD thank the European Research Council (Starting Grant 3DICE, grant agreement 336474) for fundings. VW and PG are also grateful to the CNRS program "Physique et Chimie du Milieu Interstellaire" (PCMI) for partial funding of their work.

**Table 1.** Observed molecules considered in this work[(a)]

| Organic molecules and precursors | Nitriles | Small hydrocarbons | F-bearing molecules |
|---|---|---|---|
| HCO | $CH_3CN$ | CCH | $CF^+$ |
| $H_2CO$ | $HC_3N$ | $l-C_3H$ | |
| $CH_3OH$ | $C_3N$ | $c-C_3H$ | |
| HCOOH | | $l-C_3H_2$ | |
| $CH_2CO$ | | $c-C_3H_2$ | |
| $CH_3CHO$ | | $l-C_3H^+$ | |
| $CH_3CCH$ | | | |

**Notes.** [(a)] As seen in the WHISPER survey.

**Table 3.** Visual extinction ranges at which our model reproduces the observations within their error bars.

| Species | $A_V$(mag) | | | |
|---|---|---|---|---|
| | Core region ($A_V \gtrsim 3$ mag) | | PDR region ($A_V \lesssim 3$ mag) | |
| | $4.64\times10^4$ yr | $1\times10^6$ yr | $4.64\times10^4$ yr | $1\times10^6$ yr |
| HCO | [3.9 – 4.2; $\gtrsim$ 6.5](*) | $\gtrsim$ 5.4(*) – [5.4 – 5.7; $\gtrsim$ 13] | [0.9 – 3.0](*) – [1.8 – 2.1] | [0.9 – 3.0](*) – [1.8 – 2.1] |
| $H_2CO$ | [3.3 – 4.1](*) | $\gtrsim$ 5.3(*) – [5.3 – 5.7; $\gtrsim$ 15] | [0.6 – 1.9](*) – [1.1 – 1.4] | [0.6 – 1.8](*) – [0.9 – 1.2] |
| $CH_3OH$ | $\gtrsim$ 5.4(*) | $\gtrsim$ 5.5(*) – [~ 5.5; 6.7 – 10] | [0.5 – 1.5](*) – [0.7 – 0.9] | [0.5 – 1.4](*) – [0.7 – 0.8] |
| $CH_3CCH$ | $\gtrsim$ 3.0(*)– $\gtrsim$ 3.5 | $\gtrsim$ 3.0(*) – [3.8 – 4.8; 5.9 – 6.4] | [2.3 – 3.0](*) – [2.7 – 3.0] | [2.3 – 3.0](*) – [2.7 – 3.0] |
| $H_2CCO$ | $\gtrsim$ 4(*) | [3 – 5.3; 5.7 – 6.3](*) | [2.0 – 3.0](*) | [1.9 – 3.0](*) |
| $CH_3CHO$ | $\gtrsim$ 4.2(*) | [3.7 – 5.4; 5.6 – 6.5](*) – [4.5 – 5.1] | [1.9 – 2.9](*) | [1.7 – 2.7](*) |
| HCOOH | [3.0 – 3.8] - $\gtrsim$ 4.2(*) | [3.0 – 5.2](*) – [3.0 – 3.5; 4.6 – 4.9] | [2.6 – 3.0](*) | [2.6 – 3.0](*) |
| $CH_3CN$ | – | $\gtrsim$ 5.9(*) | [2.3 – 3.0](*) – [2.6 – 2.7] | [2.3 – 3.0](*) – [2.6 – 2.7] |
| $HC_3N$ | – | [7 – 10](*) | [0.9 – 2.5](*) – [1.7 – 2.0] | [0.9 – 2.5](*) – [1.8 – 2.1] |
| $C_3N$ | – | $\gtrsim$ 6.4(*) | [0.7 – 2.7](*) – [1.0 – 2.1] | [0.7 – 2.7](*) – [1 – 2.1] |
| | $4.64\times10^2$ yr | $1\times10^6$ yr | $4.64\times10^2$ yr | $1\times10^6$ yr |
| $C_2H$ | $\gtrsim$ 6.5(*) | [3 – 6.5](*)– $\gtrsim$ 6.5(*) | [0.18 – 3.0](*) – [0.19 – 0.38] | [0.18 – 3.0](*) – [0.36 – 0.6; 1.3 – 2.0] |
| $c-C_3H$ | – | – | [0.18 – 1.1; 1.8 – 3.0](*) – [0.18 – 0.20] | [0.33 – 2.8](*) – [0.6 – 1] |
| $l-C_3H$ | – | – | [0.18 – 1.1; 1.8 – 3.0](*) – [0.18 – 0.20] | [0.33 – 3.0](*) – [0.6 – 1] |
| $c-C_3H_2$ | – | – | [0.18 – 1.2; 1.8 – 2.6](*) – [0.19 – 0.26] | [0.35 – 2.8](*) – [0.75 – 0.95] |
| $l-C_3H_2$ | – | – | [0.18 – 3](*) – [0.19 – 0.33] | [0.38 – 3.0](*) – [0.7 – 1] |
| $C_3H^+$ | $\gtrsim$ 3 | $\gtrsim$ 6.5 | [0.16 – 2.3](*) – [0.18 – 0.28; 1.5 – 1.8] | [0.14 – 1.8](*) – [0.25 – 0.6; 1.1 – 1.4] |
| $CF^+$ | $\gtrsim$ 3 | $\gtrsim$ 3 | [0.02 – 0.6] | [0.07 – 0.8] |

**Notes**:

(*) While these species are under- or over-estimated by a factor of a few in the regions of interest, if we assume an arbitrary error factor of ten on the modeled abundances, these results are still consistent with the observations.

**Table 4.** Observations and model results for fractional abundances at $10^6$ yr

| | PAH postion (*i.e.* IR-edge) | | PDR position (*i.e.* IR-peak) | | Core position | |
|---|---|---|---|---|---|---|
| | Obs. | This work | Obs. | This work | Obs. | This work |
| $A_V$ (mag): | ≃ 0.05$^{(a)}$ | 0.05 | ≃ 1.5$^{(a)}$ | 1.5 | ≃ [8 − 10]$^{(b)}$ | 8 |
| Species | | | | | | |
| O | – | 2.9(-4) | – | 2.7(-5) | – | 4.5(-9) |
| $O_2$ | – | 5.0(-12) | – | 1.4(-9) | – | 2.6(-11) |
| $O_3$ | – | 1.0(-6) | – | 5.8(-17) | – | 1.7(-15) |
| OH | – | 2.3(-9) | – | 1.7(-8) | – | 2.8(-9) |
| $H_2O$ | – | 9.9(-10) | – | 4.7(-8) | – | 1.2(-9) |
| $OH^+$ | – | 2.0(-11) | – | 8.6(-15) | – | 3.5(-16) |
| $H_2O^+$ | – | 2.1(-11) | – | 3.2(-14) | – | 3.1(-15) |
| $H_3O^+$ | – | 2.6(-11) | – | 1.6(-12) | – | 3.6(-11) |
| C | – | 5.9(-8) | – | 3.4(-6) | – | 5.6(-8) |
| CO | – | 1.0(-9) | 1.9(-7)$^{(c)}$ | 1.3(-4)$^{(†)}$ | – | 1.3(-7) |
| $C^+$ | – | 1.4(-4) | – | 7.6(-7) | – | 2.5(-10) |
| $CO^+$ | – | 2.2(-12) | ≤ 5.0(-13)$^{(c)}$ | 1.7(-13) | – | 3.0(-17) |
| $HCO^+$ | – | 8.2(-12) | 9.0(-10)$^{(c)}$ | 6.1(-11) | 3.9(-9)$^{(c)}$ | 8.6(-11) |
| $HOC^+$ | – | 1.4(-11) | 4.0(-12)$^{(c)}$ | 3.0(-11) | – | 4.2(-14) |
| **HCO** | – | 2.4(-15) | 8.4(-10)$^{(c)}$ | 1.5(-10) | < 8.0(-11)$^{(c)}$ | 2.5(-10) |
| **$H_2CO$** | – | 3.4(-13) | 2.9(-10)$^{(c)}$ | 1.9(-9) | 2.0(-10)$^{(c)}$ | 5.6(-10) |
| **$CH_3OH$** | – | 2.8(-13) | 1.3(-10)$^{(c)}$ | 4.2(-9) | 2.3(-10)$^{(c)}$ | 2.1(-10) |
| **HCOOH** | – | 2.8(-15) | 5.2(-11)$^{(c)}$ | 2.6(-15) | 1.4(-11)$^{(c)}$ | 2.8(-15) |
| **$CH_2CO$** | – | 2.0(-20) | 1.5(-10)$^{(c)}$ | 2.9(-12) | 4.9(-11)$^{(c)}$ | 6.7(-13) |
| **$CH_3CHO$** | – | 1.7(-20) | 2.4(-11)$^{(c)}$ | 1.6(-12) | 6.8(-11)$^{(c)}$ | 2.2(-13) |
| **$CH_3CCH$** | – | 3.5(-24) | 4.4(-11)$^{(c)}$ | 1.3(-13) | 3.0(-10)$^{(c)}$ | 9.2(-11) |
| $CH_3O$ | – | 1.1(-17) | – | 5.6(-12) | – | 2.0(-13) |
| $CH_3OCH_3$ | – | 1.4(-19) | – | 8.7(-17) | – | 1.7(-15) |
| $HCOOCH_3$ | – | 4.7(-27) | – | 2.4(-20) | – | 2.4(-15) |
| **$C_2H$** | [1.3 − 9.0](-8)$^{(a)}$ | 7.8(-13) | [1.9 − 5.9](-9)$^{(a)}$ | 4.6(-9) | < 1.8(-9)$^{(a)}$ | 1.3(-8) |
| **c−$C_3H$** | – | 2.7(-17) | [0.37 − 2.2](-10)$^{(a)}$ | 2.8(-10) | [0.74 − 4.1](-11)$^{(a)}$ | 1.9(-9) |
| **l−$C_3H$** | – | 5.7(-18) | [0.92 − 5.4](-11)$^{(a)}$ | 7.1(-11) | [1.3 − 7.5](-12)$^{(a)}$ | 6.0(-10) |
| **c−$C_3H_2$** | (3.0 ± 2.0)(-9)$^{(a)}$ | 1.1(-16) | [0.78 − 2.9](-10)$^{(a)}$ | 4.4(-10) | [8.4 − 2.6](-11)$^{(a)}$ | 1.5(-9) |
| **l−$C_3H_2$** | – | 2.1(-18) | [0.77 − 4.5](-11)$^{(a)}$ | 4.3(-11) | [0.83 − 4.6](-12)$^{(a)}$ | 2.2(-10) |
| **l−$C_3H^+$** | [1.2 − 7.2](-11)$^{(a)}$ | 1.8(-15) | [0.2 − 1.4](-11)$^{(a)}$ | 6.4(-13) | < 1.7(-12) $^{(a)}$ | 9.4(-14) |
| $C_4H$ | 5.2(-9)$^{(d)}$ | 3.7(-18) | 1.9(-9)$^{(d)}$ | 8.7(-11) | 3.7(-10)$^{(d)}$ | 2.2(-9) |
| $CH_4$ | – | 1.6(-13) | – | 3.5(-8) | – | 2.3(-7) |
| $C_6H$ | – | 4.2(-17) | 2.2(-11)$^{(c)}$ | 7.9(-13) | – | 2.8(-10) |
| $CH^+$ | – | 3.3(-12) | – | 1.7(-13) | – | 3.3(-15) |
| $CH_3^+$ | – | 3.7(-11) | – | 1.2(-10) | – | 5.8(-11) |
| $C_2H_4^+$ | – | 2.3(-17) | – | 6.2(-13) | – | 4.4(-11) |
| N | – | 7.9(-5) | – | 2.5(-5) | – | 1.3(-8) |
| CN | – | 1.4(-10) | – | 1.2(-7) | – | 1.8(-9) |
| NO | – | 2.9(-12) | – | 3.7(-9) | – | 1.4(-10) |
| $NH_3$ | – | 1.3(-10) | – | 7.9(-9) | – | 1.2(-8) |
| HCN | – | 4.3(-12) | – | 5.6(-9) | – | 3.4(-9) |
| HNC | – | 4.2(-12) | – | 2.5(-9) | – | 2.7(-9) |
| **$C_3N$** | – | 9.0(-20) | 5.0(-11)$^{(e)}$ | 3.6(-11) | 1.5(-11)$^{(e)}$ | 4.5(-11) |
| **$HC_3N$** | – | 6.8(-21) | 6.3(-12)$^{(c)}$ | 3.1(-12) | 7.9(-12)$^{(c)}$ | 1.2(-10) |
| $HC_5N$ | – | 1.8(-20) | – | 4.3(-14) | – | 2.2(-11) |
| **$CH_3CN$** | – | 5.5(-20) | 2.5(-10)$^{(c)}$ | 3.6(-13) | 7.9(-12)$^{(c)}$ | 1.5(-11) |
| $S^+$ | – | 3.5(-6) | – | 3.0(-6) | – | 3.6(-10) |
| CS | – | 1.4(-14) | 2.0(-9)$^{(c)}$ | 1.9(-8) | 2.9(-9)$^{(c)}$ | 7.3(-9) |
| $HCS^+$ | – | 3.0(-15) | 1.7(-11)$^{(c)}$ | 2.9(-11) | 1.2(-11)$^{(c)}$ | 2.7(-11) |
| **$CF^+$** | – | 9.8(-11) | 5.7(-10)$^{(c)}$ | 1.1(-12) | < 6.9(-11)$^{(c)}$ | 1.0(-15) |

**Notes.** Numbers in parentheses are powers of 10. Molecules in boldface correspond to those discussed in Sect. 3. $^{(a)}$ Values from Guzman et al. (2015); $^{(b)}$ Values from Pety et al. (2012) and Gerin et al. (2009); $^{(c)}$ Observed values from Guzman et al. (2014) and references therein; $^{(†)}$ Note that the CO modeled value is for the specific $A_V$ of 1.5 mag shown here, but our model also predicts a CO abundance between $10^{-7}$ and $10^{-6}$ in the $A_V$ range 0.3 − 0.5 mag ; $^{(d)}$ Observed values from Pety et al. (2005); $^{(e)}$ Observed values from Agúndez et al. (2008);

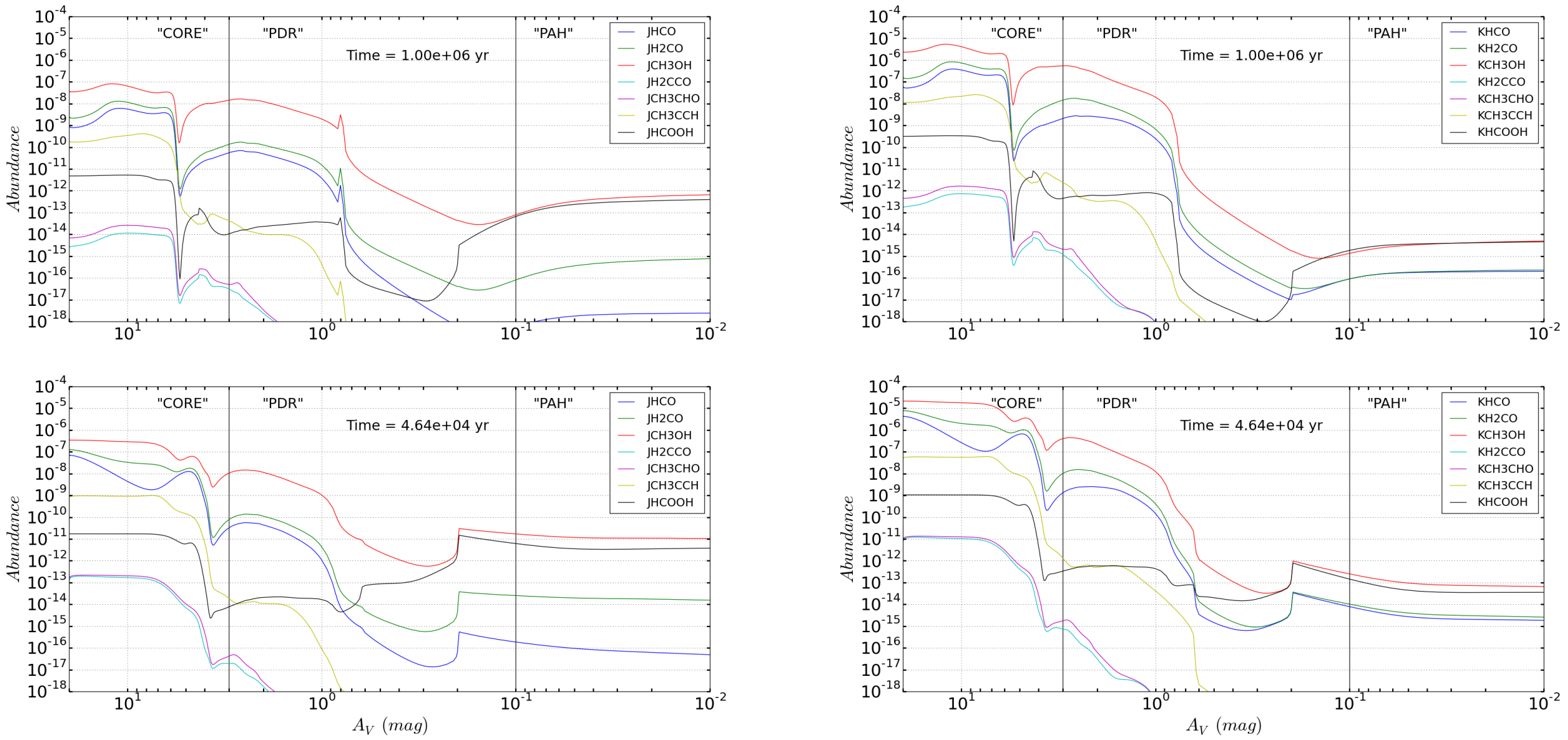

**Fig. 7.** As Fig. 2 but here the abundances of the equivalent grain mantle-surface molecules ("Jspecies") and grain bulk-mantle molecules ("Kspecies") are depicted.

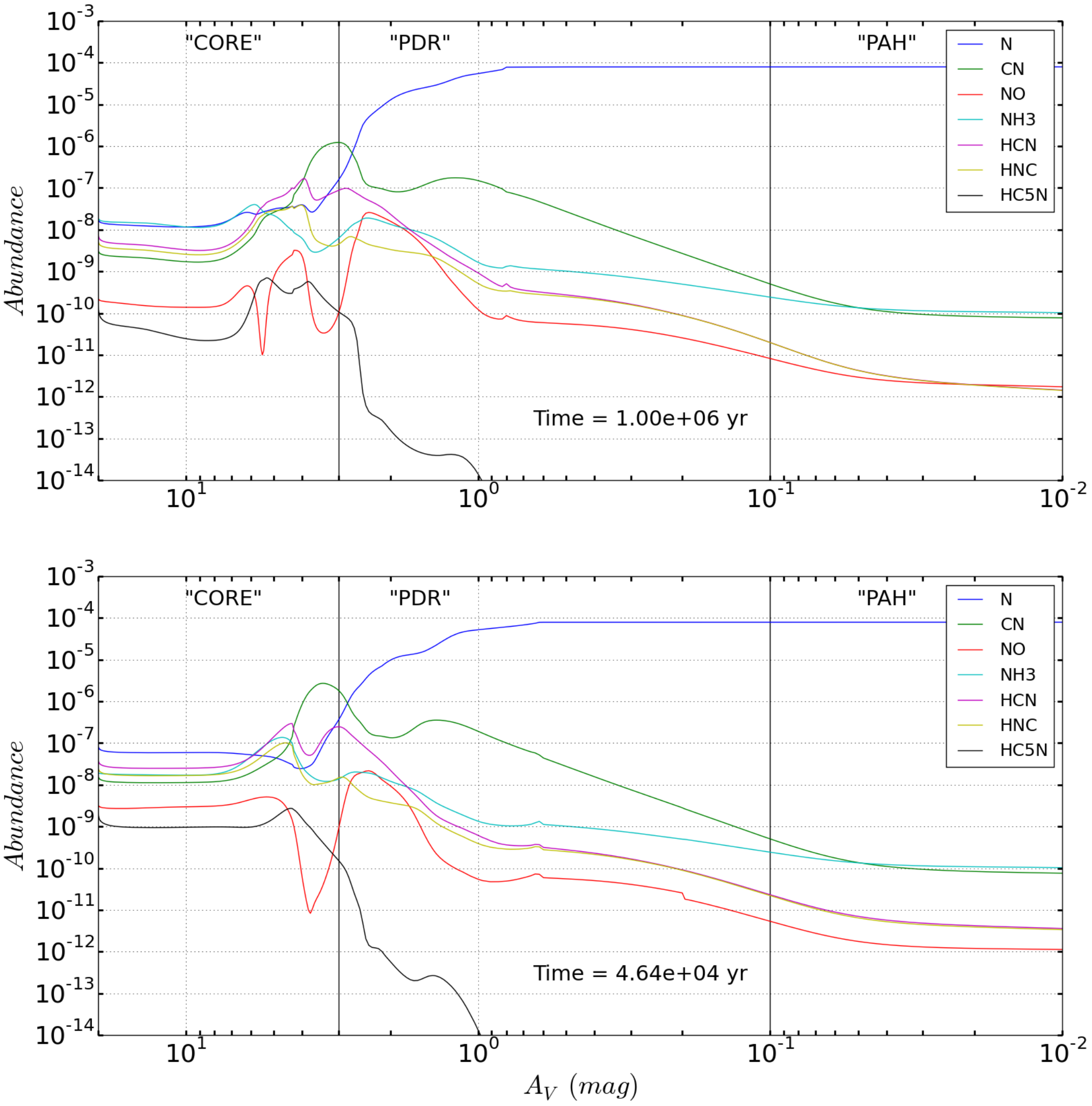

**Fig. 8.** Same as Figs. 2 - 7 but here representing the predicted abundances in the Horsehead nebula obtained with our model for selected nitrogen-containing molecules at the same typical early and late times: $4.64 \times 10^4$ yr and $1 \times 10^6$ yr.

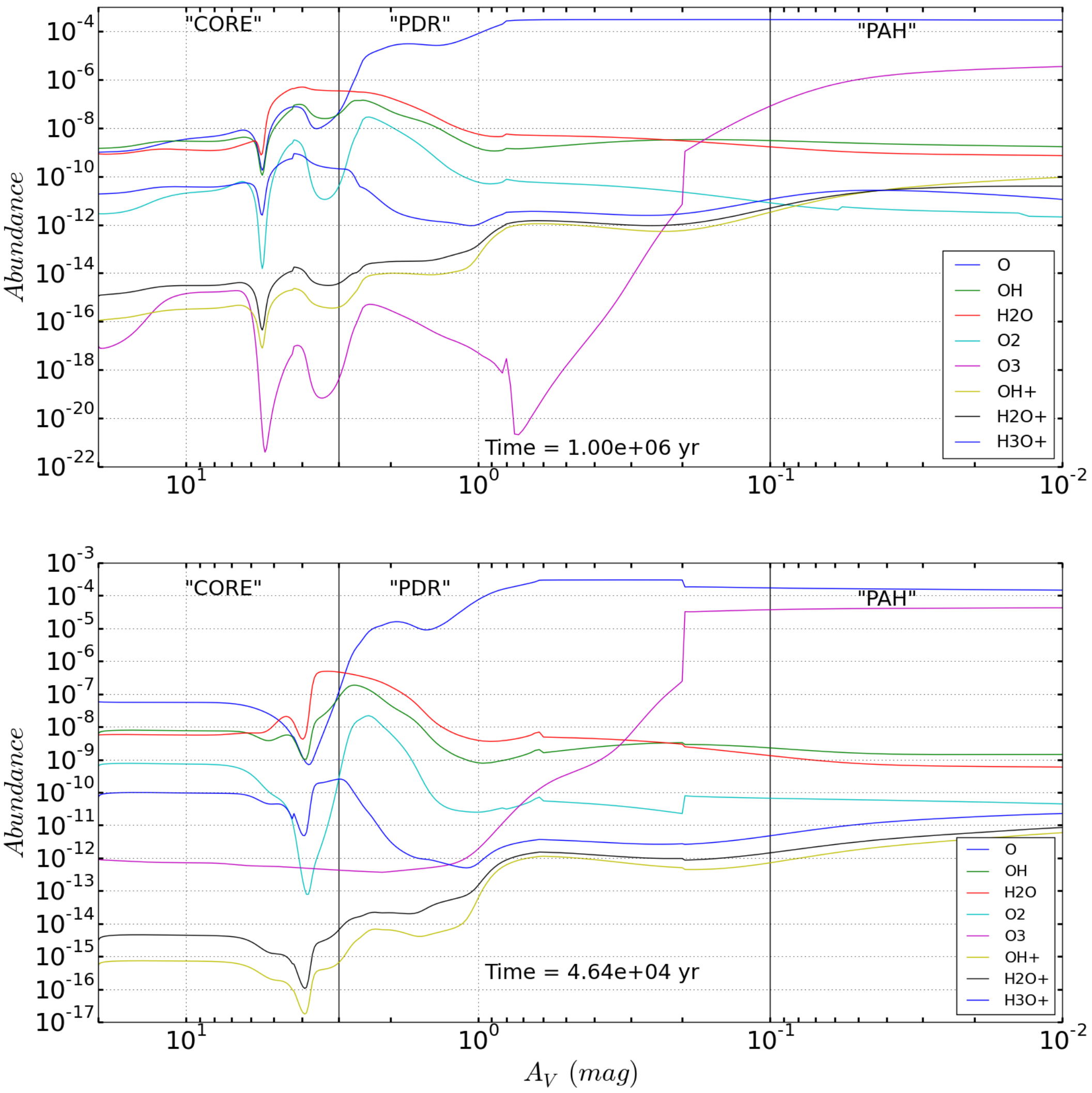

**Fig. 9.** As Fig. 8 but here representing the predicted abundances in the Horsehead nebula obtained with our model for selected oxygen-containing molecules at the same typical early and late times.

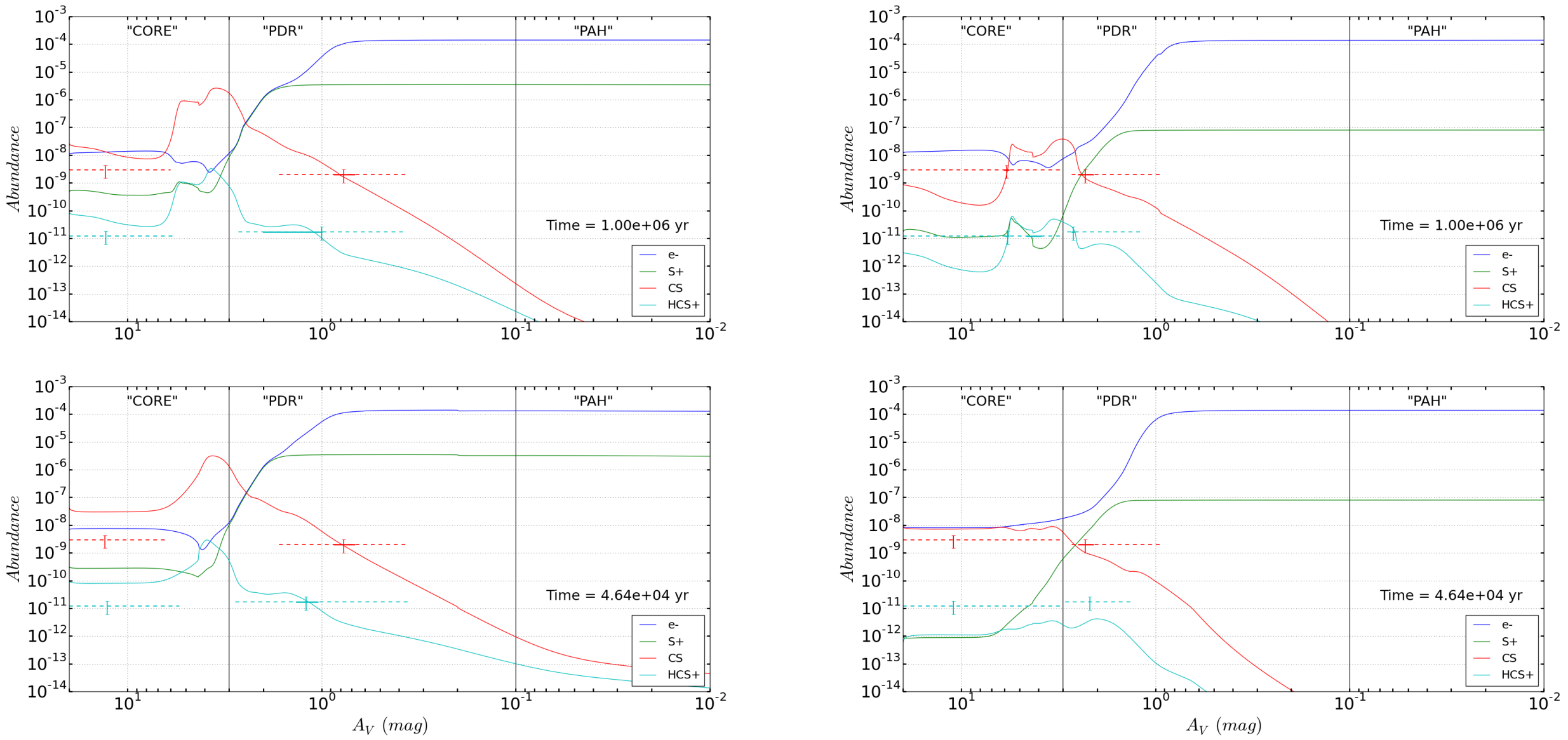

**Fig. 10.** Same as Figure 8 and 9 but here representing the predicted abundances for selected sulfur-bearing species for two different elemental amount of sulfur: a high elemental abundance of $3.5 \times 10^{-6}$ in the left panels and low elemental abundance of $8.0 \times 10^{-8}$ in the right panels (see Sect. 4.5). The error bars represent the observational values from Guzman et al. (2014).